\documentclass[12pt]{article}
\usepackage{natbib}
\usepackage{color}
\usepackage{latexsym}
\usepackage{amsthm}
\usepackage{multirow}
\usepackage{float}
\usepackage{wrapfig}
\usepackage[latin1]{inputenc}
\usepackage[T1]{fontenc}
\usepackage{amsmath,amsfonts,amssymb,bm,dsfont}
\usepackage{pslatex}
\usepackage{graphicx,color}
\usepackage{fancyhdr}
\usepackage{multicol}

\makeatletter
\renewcommand\@biblabel[1]{}

\makeatother
\usepackage[displaymath]{lineno}
\usepackage[latin1]{inputenc}
\usepackage[T1]{fontenc}
\usepackage{graphicx}
\usepackage{times}
\usepackage{amsmath,amsfonts,amssymb,bm}
\usepackage[english]{babel}
\usepackage{amsbsy}

\usepackage{subfigure}
\usepackage{multirow}
\usepackage{multicol}

\newenvironment{remark}[1][Remark]{\begin{trivlist}
\item[\hskip \labelsep {\bfseries #1}]}{\end{trivlist}}
\theoremstyle{definition}
\newtheorem{cond}{Condition}
 \newcommand*\patchAmsMathEnvironmentForLineno[1]{%
  \expandafter\let\csname old#1\expandafter\endcsname\csname #1\endcsname
  \expandafter\let\csname oldend#1\expandafter\endcsname\csname end#1\endcsname
  \renewenvironment{#1}%
     {\linenomath\csname old#1\endcsname}%
     {\csname oldend#1\endcsname\endlinenomath}}%
\newcommand*\patchBothAmsMathEnvironmentsForLineno[1]{%
  \patchAmsMathEnvironmentForLineno{#1}%
  \patchAmsMathEnvironmentForLineno{#1*}}%
\AtBeginDocument{%
\patchBothAmsMathEnvironmentsForLineno{equation}%
\patchBothAmsMathEnvironmentsForLineno{align}%
\patchBothAmsMathEnvironmentsForLineno{flalign}%
\patchBothAmsMathEnvironmentsForLineno{alignat}%
\patchBothAmsMathEnvironmentsForLineno{gather}%
\patchBothAmsMathEnvironmentsForLineno{multline}%
}

\begin{document}

\begin{center}

\Large {\bf The use of spatial information in entropy measures}\par

\bigskip

\normalsize{Linda Altieri, Daniela Cocchi, Giulia Roli}\\ \small{Department of Statistics, University of Bologna, via Belle Arti, 41, 40126, Bologna, Italy.} \par

\bigskip

%

\end{center}

\begin{quotation}
\noindent {\it Abstract:}
The concept of entropy, firstly introduced in information theory, rapidly became popular in many applied sciences via Shannon's formula to measure the degree of heterogeneity among observations. A rather recent research field aims at accounting for space in entropy measures, as a generalization when the spatial location of occurrences ought to be accounted for. The main limit of these developments is that all indices are computed conditional on a chosen distance. This work follows and extends the route for including spatial components in entropy measures. Starting from the probabilistic properties of Shannon's entropy for categorical variables, it investigates the characteristics of the quantities known as residual entropy and mutual information, when space is included as a second dimension. This way, the proposal of entropy measures based on univariate distributions is extended to the consideration of bivariate distributions, in a setting where the probabilistic meaning of all components is well defined. As a direct consequence, a spatial entropy measure satisfying the additivity property is obtained, as global residual entropy is a sum of partial entropies based on different distance classes. Moreover, the quantity known as mutual information measures the information brought by the inclusion of space, and also has the property of additivity. A thorough comparative study illustrates the superiority of the proposed indices.
\bigskip
\end{quotation}

\begin{quotation}
\bigskip
\end{quotation}

\section{Introduction}

When a set of units can be assigned to a finite number of categories of a study variable, a popular way of assessing heterogeneity is to compute entropy. The concept of entropy has been firstly introduced in information theory to evaluate the degree of heterogeneity in signal processing. The seminal work by Shannon (\citeyear{shan}) provided the basics to define entropy, and Shannon's formula of entropy rapidly became popular in many applied sciences, e.g. ecology and geography \citep{patiltaillie, hoeting, frosini, cobbold}. The reasons for the success of this measure are two-fold. On one hand, entropy is a measure of diversity that only explicitly considers the number of categories of the study variable and their probabilities of occurrence; thus, it can be employed in a wide range of applications, even when qualitative variables are involved. On the other hand, entropy summarizes and captures several aspects that are differently denoted according to the specific target: heterogeneity, information, surprise, diversity, uncertainty, contagion are all concepts strongly related to entropy. Information theory also investigates the relationship across two variables under a probabilistic framework to form bivariate entropy measures that, despite their interesting properties (\citeauthor{renyi}, \citeyear{renyi}, \citeauthor{coverthomas}, \citeyear{coverthomas}), are as yet not deeply explored.

A relatively recent research field aims at accounting for space in entropy measures, as a natural generalization when the spatial location of the occurrences is available and relevant to the purpose of the analysis. Spatial data are georeferenced over spatial units which may be points or areas; this work deals with areal data, but most measures are applicable to point data as well. Spatial entropy must not be confused with spatial correlation, a measure that also identifies the type of spatial association, either positive or negative. Indeed, entropy measures heterogeneity, irrespective of the type of association across the outcomes of the study variable. Several solutions to account for space in entropy measures have been proposed in geography, ecology and landscape studies, from the papers by Batty (\citeyear{batty74}, \citeyear{batty76}) to recent works \citep{batty10, cobbold, batty14, leibovici14}. The underlying idea is to include spatial information into the computation of entropy for capturing the role of space in determining the occurrences of interest. 

Being entropy an expectation, care has to be devoted at defining the entity to which a probability is assigned, since the sum of all probabilities has to be 1. In most entropy measures, this entity is the $i$-th category of a variable $X$, say $x_i$, with $i=1, \dots, I$, and the probability of this occurrence is denoted by $p(x_i)$. The statistical properties of Shannon's entropy are usually assessed under this definition. When considering Batty's spatial entropy, the categories of the study variables are defined according to the $g=1, \dots, G$ zones that partition a territory, and the probability $p_g$ represents the probability of occurrence of the phenomenon of interest over zone $g$. Other approaches to spatial entropy (\citeauthor{oneill}, \citeyear{oneill}, \citeauthor{leibovici09}, \citeyear{leibovici09}) are based on a transformation of the study variable, aimed at including space, with its probability distribution being employed in Shannon's formula. 

The present work proposes spatial entropy measures which exploit the full probabilistic framework provided in the information theory field and are based on both univariate and bivariate distributions. Our approach leads to coherently defined entropy measures that are able to discern and quantify the role of space in determining the outcomes of the variable of interest, that needs to be suitably defined. Indeed, the entropy of such variable can be decomposed into spatial mutual information, i.e. the entropy due to space, and spatial residual entropy, i.e. the remaining information brought by the variable itself once space has been considered. Furthermore, the present proposal solves the problem of preserving additivity in constructing entropy measures, that was pioneeringly tackled by \citeauthor{theil} (\citeyear{theil}), allowing for partial and global syntheses. The topic of additivity has been thoroughly faced in a number of papers \citep{anselin, karlstrom, entrcontinuo}, but without exploiting the properties of entropy computed on bivariate distributions.

The remainder of the paper is organized as follows. In Section \ref{sec:prop} some properties of classical entropy are highlighted; Section \ref{sec:reviewspace} reviews popular spatial entropy measures. In Section \ref{sec:nostra} an innovative way to deal with space in entropy is proposed, which is thoroughly assessed in Section \ref{sec:sim}. Section \ref{sec:disc} discusses main results and concludes the paper.

\section{Statistical properties of Shannon's entropy}
\label{sec:prop}

Information theory provides a complete probabilistic framework to properly define entropy measures in the field of signal processing. Its original aim is to quantify the amount of information related to a message when it is transmitted by a channel that can be noisy \citep{coverthomas, stone}. The message is treated as a discrete random variable, $X$, which can assume a set of different values, $x_i$, $i=1,\dots,I$ where $I$ is the number of possible outcomes. The term information is associated to the concepts of surprise and uncertainty: the greater the surprise (and, thus, the uncertainty) in observing a value $X=x_i$, the greater the information it contains. The amount of surprise about an outcome value $x_i$ increases as its probability decreases. In this spirit, \citeauthor{shan} (\citeyear{shan}) introduced the random variable $I(p_X)$ called information function, where $p_X=(p(x_1),\dots,p(x_{I}))'$ is the univariate probability mass function (pmf) of $X$. The information function takes values $I(p(x_i))=\log(1/p(x_i))$. It measures the amount of information contained in an outcome which has probability $p(x_i)$, without any focus on the value of the outcome itself. In information theory, the logarithm has base $2$ to quantify information in bits, but this point is irrelevant since entropy properties are invariant with respect to the choice of the base.

The entropy, also known as Shannon's entropy \citep{shan}, of a variable $X$ with $I$ possible outcomes is then defined as the expected value of the information function
\begin{equation}
H(X)=E[I(p_X)]=\sum_{i=1}^{I} p(x_i)\log\left(\frac{1}{p(x_i)}\right).
\label{eq:shann}
\end{equation}
Being an expected value, it measures the average amount of information brought by the realizations of $X$ as generated by the pmf $p_X$. When entropy is high, no precise information is available about the next realization, therefore the amount of information it brings is high. On the other hand, if one is fairly sure about the next observation, its occurrence does not carry much information, and the entropy is low.
The probabilistic properties of entropy are often left apart in the applied literature and entropy is commonly seen as a heterogeneity index, which can be computed without the value of the study variable for the different categories. Entropy $H(X)$ ranges in $[0,\log(I)]$, i.e. it is nonnegative and its maximum depends on the number of categories of $X$. The maximum value of entropy is achieved when $X$ is uniformly distributed, while the minimum is only reached in the extreme case of certainty about the variable outcome. In order to let entropy vary between $0$ and $1$, a suitable positive constant $B$, equal to the inverse of the entropy maximum value, is often introduced to obtain the normalized version of the index:
\begin{equation}H_{norm}(X)=B\cdot H(X)=\frac{H(X)}{\log(I)}.\label{eq:relH}
\end{equation}

When two pmfs for $X$ are competing, say $p_X$ and $q_X$, being $q_X$ the reference distribution, a measure of distance between the two is defined in terms of entropy. This quantity is called Kullback-Leibler distance, or relative entropy:
\begin{equation}D_{KL}(p_X || q_X)= E\left[I\left(\frac{q_X}{p_X}\right)\right]=\sum_{i=1}^{I} p(x_i)\log\left(\frac{p(x_i)}{q(x_i)}\right) \label{eq:kullback}
\end{equation}
where the weights of the expectation come from the pmf $p_X$, i.e. the distribution in the denominator of the information function. Being a distance measure, any Kullback-Leibler distance is non-negative.
\begin{remark} When $q_X$ is the uniform distribution $U_X$, expression (\ref{eq:kullback}) is the difference between the maximum value of $H(X)$ and $H(X)$ itself
\begin{equation}D_{KL}(p_X || U_X)= \log(I)-H(X). \label{eq:kullback_unif}
\end{equation} \end{remark} 

When a noisy channel is considered in information theory, a crucial point is represented by the importance of discerning the amount of information related to $X$ from the noise. In such cases, a further message, the original non-noised one, is introduced as a second discrete random variable $Y$ with $j=1, \ldots, J$ potential outcomes $y_j$. A pmf $p_Y$ is associated to the variable $Y$, and the marginal entropy $H(Y)$ can be similarly computed. This suggests to adopt a bivariate perspective; various kinds of expectations with different properties can be thus derived with reference to a joint pmf $p_{XY}$ \citep{coverthomas, stone}.

A crucial quantity is represented by the expectation known as mutual information of $X$ and $Y$, defined as
\begin{equation}
MI(X,Y)=E\left[I\left(\frac{p_Xp_Y}{p_{XY}}\right)\right]=\sum_{i=1}^{I}\sum_{j=1}^{J}p(x_i,y_j)\log{\left(\frac{p(x_i,y_j)}{p(x_i)p(y_j)}\right)}.
\label{eq:mutual}
\end{equation}
Expression (\ref{eq:mutual}) is a Kullback-Leibler distance $D_{KL}(p_{XY}||p_Xp_Y)$, where the reference joint pmf is the independence distribution of $X$ and $Y$, $p_Xp_Y$; the terms $I\left(p(x_i)p(y_j)/p(x_i,y_j)\right)$ of the information function are farther from $0$ as the association $(i,j)$ moves away from independence. When the two variables are independent, i.e. $p_{XY}=p_Xp_Y$, the mutual information is null, since, for all $i$ and $j$, $\log{\left(p(x_i,y_j)/p(x_i)p(y_j)\right)}=0$.
Mutual information measures the association of the two messages, i.e. the amount of information of $X$ due to $Y$ (or vice-versa, as the measure is symmetric), thus removing the noise effect.

Expression (\ref{eq:mutual}) can be also seen as a measure with the same structure as (\ref{eq:shann}):
\begin{equation}
MI(X,Y)=\sum_{i=1}^{I}p(x_i)\sum_{j=1}^{J}p(y_j|x_i)\log{\left(\frac{p(y_j|x_i)}{p(y_j)}\right)},
\label{eq:mutual_sommai}
\end{equation}
where, for each $i$, the information function in (\ref{eq:shann}) is replaced by a Kullback-Leibler distance $D_{KL}(p_{Y|x_i}||p_Y)$. This distance assesses how much, on average, each value of the conditional distribution $p_{Y|x_i}$ differs from the marginal $p_Y$, i.e. from independence. 

\begin{remark} Mutual information is both a Kullback-Leibler distance on a joint pmf and a weighted sum of Kullback-Leibler distances on univariate pmfs; being a symmetric measure, the decomposition holds in both directions, so that it is also a weighted sum of $D_{KL}(p_{X|y_j}||p_X)$.
\end{remark}

A further important measure of entropy, known as conditional entropy, involves the joint pmf $p_{XY}$ as (\ref{eq:mutual}) and is defined as
\begin{equation}
\begin{split}
H(X)_Y&=E\left[H(X|y_j)\right]=\sum_{j=1}^{J}p(y_j)H(X|y_j)\\
&=E\left[E\left(I\left(p_{X|y_j}\right)\right)\right]=\sum_{j=1}^{J}p(y_j)\sum_{i=1}^{I}p(x_i|y_j)\log{\left(\frac{1}{p(x_i|y_j)}\right)}\\ &=\sum_{i=1}^{I}\sum_{j=1}^{J}p(x_i,y_j)\log{\left(\frac{1}{p(x_i|y_j)}\right)}.
\end{split}
\label{eq:residYX}
\end{equation}
In information theory, this quantity is also called residual or noise entropy, as it expresses the residual amount of information brought by $X$ once the influence of the non-noised $Y$ has been removed. The components $H(X|y_j)=E\left[I\left(p_{X|y_j}\right)\right]$ of (\ref{eq:residYX}) are entropies. For this reason, $H(X)_Y$ enjoys the additive property, i.e. (\ref{eq:residYX}) is an example of the law of iterated expectations, being the expectation of a conditional expectation, while marginal entropy (\ref{eq:shann}) is not.

Residual entropy (\ref{eq:residYX}), like (\ref{eq:mutual}), maintains the same structure as (\ref{eq:shann}):
\begin{equation}
H(X)_Y=\sum_{i=1}^{I}p(x_i)\sum_{j=1}^{J}p(y_j|x_i)\log{\left(\frac{1}{p(x_i|y_j)}\right)}
\label{eq:rrresid}
\end{equation}
where, analogously to what observed in (\ref{eq:mutual_sommai}), the information function in (\ref{eq:shann}) is replaced by a more complex synthesis. 

If $Y$ partially explains $X$, the entropy of $X$ should be lower when $Y$ is taken into account. Indeed, it has been shown \citep{coverthomas} that
\begin{equation}
MI(X,Y)=H(X)-H(X)_Y=H(Y)-H(Y)_X,
\label{mutrul}
\end{equation}
that is, marginal entropy is the sum of mutual information and residual entropy. Since the concept of mutual information is symmetric, both equalities in (\ref{mutrul}) hold, where residual entropy $H(Y)_X$ can be defined analogously to (\ref{eq:residYX}). 
When independence occurs, $H(X)_Y=H(X)$, since knowing $Y$ does not reduce the uncertainty related to (i.e. the amount of information carried by) a realization of $X$. On the contrary, if there were a perfect relation between $X$ and $Y$, then $H(X)=MI(X,Y)$ and the residual entropy would be zero. In non-extreme situations, any additive term in (\ref{eq:mutual}) can be explored to check what simultaneous realizations of $X$ and $Y$ are farther away from independence; the same elementwise investigation can be performed for any of the $J$ random components $H(X|y_j)$ in (\ref{eq:residYX}).

Another important quantity is the joint entropy $H(X,Y)$, which is the equivalent of (\ref{eq:shann}) when a joint pmf $p_{XY}$ is considered:
\begin{equation}
H(X,Y)=E\left[I\left(p_{XY}\right)\right]=\sum_{i=1}^{I}\sum_{j=1}^{J}p(x_i,y_j)\log{\left(\frac{1}{p(x_i,y_j)}\right)}
\label{eq:joint}
\end{equation}
and expresses the total amount of information given by a simultaneous realization of $X$ and $Y$. The term 'joint' does not take the usual statistical meaning because it is not a measure of association, rather a total measure of the entropy of $X$ and $Y$ together. Therefore, $H(X,Y)$ is also called total entropy in the information theory language.

The following symmetric property holds \citep{coverthomas}:
\begin{equation}
H(X,Y)=H(X)+H(Y)-MI(X,Y)\label{eq:rule3}
\end{equation}
with $H(X,Y)=H(X)+H(Y)$ in the case of independence.
\begin{remark} An interesting special case of independence occurs when $p_{XY}$ is uniform, i.e. $U_{XY}$. In this situation, not only $MI(X,Y)=0$, $H(X)=H(X)_Y$ and $H(Y)=H(Y)_X$ but, in addition, the marginal entropies $H(X)$ and $H(Y)$ reach their theoretical maxima, $\log(I)$ and $\log(J)$, with the consequence that the total entropy is $H(X,Y)=\log(I)+\log(J)$. Indeed, this case describes the situation with the maximum uncertainty about the possible outcomes, i.e. with the highest marginal, residual and total entropy.\end{remark}

A further relationship between entropy indices, similar to (\ref{mutrul}), is
\begin{equation}
H(X,Y)=H(Y)_X+H(X)=H(X)_Y+H(Y)\label{eq:rule5}.
\end{equation}
This equation states that the total entropy can be computed by summing the residual entropy and the marginal entropy of the variable assumed as conditioning \citep{coverthomas}.

Relationships (\ref{mutrul}) and (\ref{eq:rule5}) involve the three fundamental entropy measures; all are expectations of different random variables, i.e. the information functions, weighted by different probability distributions, either univariate or bivariate.

When the study variables are continuous, entropy measures cannot be simply generalized by Shannon's definition. Switching from a probability mass function (pmf) to a probability density function (pdf) unfortunately generates a measure of entropy which tends to infinity. A commonly adopted solution \citep{renyi} only considers the finite part of the entropy measure, called differential entropy, that constitutes the basis for defining Batty's spatial entropy (see Section \ref{sec:spatent}).

\section{Univariate approaches to the use of spatial information in entropy measures}
\label{sec:reviewspace}

Several fields of application of entropy indices, such as geography, ecology and landscape studies, usually deal with spatial data, i.e. data collected over an area, from now on called observation window, where the spatial location of the occurrences is relevant to the analysis. A major drawback of using classical entropy indices in such studies is that they only employ the probability of occurrence of a category without considering the spatial distribution of such occurrence. Hence, datasets with identical pmf but very different spatial configurations share the same marginal entropy, say $H(X)$: the same $H(X)$ occurs in the two cases of strong spatial association and complete random pattern, in spite of the opposite spatial configurations, since the only element entering \ref{eq:shann} is the pmf of $X$.
For this reason, a concern when computing entropy measures is the introduction of some spatial information into the formulae for capturing the distribution over space, making use, sometimes implicitly, of the concept of neighbourhood.

The notion of neighbourhood is a basic concept of spatial statistics, linked to the idea that occurrences at certain locations may be influenced, in a positive or negative sense, by what happens at surrounding locations, i.e. their neighbours. The spatial extent of the influence, i.e. the choice of the neighbourhood system, is usually fixed exogenously, prior to the analysis. The system can be represented by a graph \citep{bondy}, where each location is a vertex and neighbouring locations are connected by edges. The simplest way of representing a neighbourhood system is via an adjacency matrix, i.e. a square matrix whose elements indicate whether pairs of vertices are adjacent or not in the graph. For a simple graph representing $G$ spatial units, the adjacency matrix $A=\{a_{gg'}\}_{g,g'=1,\dots,G}$ is a square $G\times G$ matrix such that $a_{gg'}=1$ when there is an edge from vertex $g$ to vertex $g'$, and $a_{gg'}=0$ otherwise; in other words, $a_{gg'}=1$ if $g' \in \mathcal{N}(g)$, the neighbourhood of area $g$. Its diagonal elements are all zero by default. Often, a row-standardized version of $A$ is used, i.e. all $G$ rows are constrained to sum to 1. Note that the spatial units may be points, defined via coordinate pairs, or areas, identified via a representative coordinate pair, such as the area centroid. Coordinates are used to measure distances and define what spatial units are neighbours. Areal units are seen as blocks, where a single value of the study variable is observed and the neighbourhood system is built.

The idea of neighbourhood underlies a number of proposals in research fields that actively contribute to the definition of spatial measures. For instance, \citeauthor{illian} (\citeyear{illian}) propose a generalized and flexible measure of spatial biodiversity based on graphs. \citeauthor{cobbold} (\citeyear{cobbold}) present an approach to the idea of neighbourhood which measures the logical similarity between pairs of species, and is therefore more extended than the similarity between spatial units. Under this perspective, they consider a new family of diversity measures, including some entropy-based indices as special cases, in order to account for both abundances of species and differences between them.

Over the past thirty years, several works developed the idea of spatial entropy measures based on an idea of neighbourhood; they can be ascribed to two main univariate approaches. The first one (\citeauthor{batty76} \citeyear{batty74, batty76, batty10}, \citeauthor{karlstrom} \citeyear{karlstrom}),  presented in Section \ref{sec:spatent}, starts from the theory of spatial statistics but pays the price of discarding the categories of $X$.  In particular, \citeauthor{karlstrom} (\citeyear{karlstrom}) aim at building an additive measure following the idea of Local Indices of Spatial Association (LISA) proposed by \cite{anselin}. The second approach computes entropy measures based on a transformation of the study variable $X$, accounting for the distance at which specific associations of $X$ outcomes occur. The resulting measures are not additive (therefore not decomposable) and are only able to consider one distance range at a time. They are outlined in Section \ref{sec:Z}. All the above proposals, except for Batty's work, refer to the concept of neighbourhood by setting an adjacency matrix.

\subsection{Towards an additive spatial entropy}
\label{sec:spatent}

\subsubsection{Batty's spatial entropy}

One appreciable attempt to include spatial information into Shannon's entropy starts from a reformulation of (\ref{eq:shann}). The categorical variable $X$ is recoded into $I$ dummy variables, each identifying the occurrence of a specific category of $X$. The probability of success of each dummy variable, i.e. 'occurrence of the $i$-th category of $X$', is labelled as $p_i$, and, for each $i$, $1-p_i=\sum_{i' \ne i} p_{i'}$. This means that each non-occurrence of the $i$-th category of $X$ implies the occurrence of another category. As a consequence, $\sum_i p_i=1$, since the collection of occurrences constitutes a partition of the certain event. Due to the way the $I$ 'occurrence/non-occurrence' variables are defined, $p_i=p(x_i)$. Therefore, Shannon's entropy of the variable $X$ may be expressed as $H(X)=\sum_{i=1}^{I} p_i \log(1/p_i)$. 

This approach is taken by Batty (\citeyear{batty74, batty76}) to define a spatial entropy which extends Theil's work (\citeyear{theil}). In a spatial context, a phenomenon of interest $F$ occurs over an observation window of size $T$ partitioned into $G$ areas of size $T_g$, with $\sum_{g=1}^G T_g=T$. This defines $G$ dummy variables identifying the occurrence of $F$ over a generic area $g$, $g=1, \dots, G$.
Given that $F$ occurs somewhere over the window, its occurrence in zone $g$ takes place with probability $p_g$, where again $1-p_g=\sum_{g' \ne g} p_{g'}$ and $\sum_g p_g=1$. Since the collection of $p_g$ meets the criteria for being a pmf, it is possible to define the phenomenon pmf over the window $p_F=\left(p_1, \dots, p_g, \dots, p_G\right)'$. When $p_g$ is divided by the area size $T_g$, the phenomenon intensity is obtained: $\lambda_g=p_g/T_g$, assumed constant within each area $g$.

Shannon's entropy of $F$ may be written as
\begin{equation}
H(F)=E\left[I\left(p_F\right)\right]=\sum_{g=1}^G p_g \log \left(\frac{1}{p_g}\right)=\sum_{g=1}^G \lambda_gT_g \log \left(\frac{1}{\lambda_g}\right)+\sum_{g=1}^G \lambda_gT_g \log \left(\frac{1}{T_g}\right).
\end{equation}
\cite{batty76} shows that the first term on the right hand side of the formula converges to the continuous version of Shannon's entropy \citep{renyi}, namely the differential entropy, as the area size $T_g$ tends to zero. The second term is discarded and the differential entropy is rewritten in terms of $p_g$, giving Batty's spatial entropy
\begin{equation}
H_B(F)=\sum_{g=1}^G p_g \log \left(\frac{T_g}{p_g}\right).
\label{eq:spaten}
\end{equation}
It expresses the average surprise (or amount of information) brought by the occurrence of $F$ in an area $g$, and aims at computing a spatial version of Shannon's entropy. Shannon's entropy is high when the $I$ categories of $X$ are equally represented over a (non spatial) data collection, while Batty's entropy is high when the phenomenon of interest $F$ is equally intense over the $G$ areas partitioning the observation window ($\lambda_g=\lambda$ for all $g$). Batty's entropy includes a multiplicative component $T_g$ related to space in the information function that accounts for unequal space partition.

Batty's entropy $H_B(F)$ reaches a minimum value equal to $\log(T_{g^*})$ when $p_{g^*}=1$ and $p_g=0$ for all $g\ne g^*$, with $g^*$ denoting the area with the smallest size. The maximum value of Batty's entropy is $\log(T)$, reached when the intensity of $F$ is the same over all areas, i.e. $\lambda_g=1/T$ for all $g$.
This maximum value does not depend on the area partition, nor on the nature of the phenomenon of interest $F$ (discrete or continuous), but only on the size of the observation window. 
When $T_g=1$ for each $g$, $H_B(F)$ is a Shannon's entropy of $F$ equivalent to (\ref{eq:shann}), and the index ranges accordingly in $[0,\log(G)]$.

\subsubsection{A LISA version of Batty's spatial entropy}
\label{sec:kc}

A challenging attempt to introduce additive properties and to include neighbourhood in Batty's entropy index (\ref{eq:spaten}) is due to \citeauthor{karlstrom} (\citeyear{karlstrom}), following the LISA theory.

Local Indices of Spatial Association (LISA, see \citeauthor{anselin}, \citeyear{anselin}, for an extensive introduction and \citeauthor{cliff}, \citeyear{cliff},\ for the popular Moran's $I$ example of a LISA measure) are descriptive measures of a spatial dataset that satisfy the following conditions.
\begin{cond}\label{cond1}
For every spatial unit $g$ within the observation window, a LISA index measures the degree of spatial clustering/aggregation around that location; it can be viewed as a local index because it is a function of the study variable at unit $g$ and at neighbour units $g'$. In the context of Section \ref{sec:spatent}, the local index may be defined as $L_g=f(\lambda_g, \lambda_{g'\in \mathcal{N}(g)})$, where $\mathcal{N}(g)$ is the neighbourhood of $g$.
\end{cond}
\begin{cond}\label{cond2}
The sum of the indices at all spatial units $g$ is proportional to the overall index in the observation window: $\alpha\sum_g L_g=L$ where $L$ is a global version of the index. This is the desirable additivity property of local spatial measures.
\end{cond}

\citeauthor{karlstrom}'s entropy index $H_{KC}(F)$ starts by weighting the probability of occurrence of the phenomenon of interest $F$ in a given spatial unit $g$, $p_g$, with its neighbouring values:
\begin{equation}
\widetilde{p}_g=\sum_{g'=1}^G a_{gg'}p_{g'}
\label{eq:karl_ptilde}
\end{equation} 
where $a_{gg'}$ is the element of the row-standardised $G\times G$ adjacency matrix $A$, which selects the neighbouring areas and the associated probabilities $p_{g'}$. Then, an information function is defined, fixing $T_g=1$, as $I(\widetilde{p}_g)=\log\left(1/\widetilde{p}_g\right)$. When all weights are equal, i.e. $a_{gg'}=1/|\mathcal{N}(g)|$ for $g' \in \mathcal{N}(g)$ where $|\mathcal{N}(g)|$ is the cardinality of $\mathcal{N}(g)$, then an average of the $p_{g'}$ is obtained: $\sum_{g'=1}^G a_{gg'}p_{g'}=\sum_{g'\in \mathcal{N}(g)}p_{g'}/|\mathcal{N}(g)|$. In this proposal, the elements on the diagonal of the adjacency matrix $A$ are non-zero, i.e. each area neighbours itself and enters the computation of $I(\widetilde{p}_g)$. Thus, \citeauthor{karlstrom}'s entropy index is
\begin{equation}
H_{KC}(F)=E\left[I\left(\widetilde{p}_g\right)\right]=\sum_{g=1}^G p_g\log\left(\frac{1}{\widetilde{p}_g}\right).
\label{eq:karl}
\end{equation}
It can be shown that the maximum of $H_{KC}(F)$ does not depend on the choice of the neighbourhood and is $\log(G)$. As the neighbourhood reduces until vanishing, i.e. as $A$ becomes the identity matrix, $H_{KC}(F)$ coincides with Batty's spatial entropy (\ref{eq:spaten}) in the case of all $T_g=1$.  

The local measure that satisfies LISA Condition \ref{cond1} is $L_g=p_gI(\widetilde{p}_g)$. 
The sum of local measures $L_g$ forms the global index (\ref{eq:karl}), preserving the LISA property of additivity, Condition \ref{cond2}, with $\alpha=1$ as proportionality constant. The main disadvantage of the local components is that they are not expectations, therefore they are not entropy measures.

\subsection{Spatial entropies based on a transformation of the study variable}
\label{sec:Z}

A second way to build a spatial entropy measure consists in defining a new categorical variable $Z$, where each realization identifies groups of occurrences of $X$ over space, namely co-occurrences. The definition of $Z$ underlies a choice for $m$, the degree of co-occurrences of $X$ (couples, triples and so on) and an assumption on whether to preserve the order of realizations over space. Preserving the order means, for example, that a co-occurrence $(x_i,x_j)$ of degree $m=2$ is different from $(x_j, x_i)$. Then, for a generic degree $m$ and $I$ categories of $X$, the new variable $Z$ has $R^o_m=I^m$ categories; should the order not be preserved, $R^{no}_m=\binom{I+m-1}{m}$. Once $Z$ is defined, its pmf is $p_{Z}=(p(z_1), \dots, p(z_{R_m}))'$, where $p(z_r)$ is the probability of observing the $r$-th co-occurrence of $X$ over different spatial units, and $R_m$ may be alternatively $R^o_m$ or $R^{no}_m$. The pmf $p_Z$ can be used to compute Shannon's entropy (\ref{eq:shann}) of $Z$, $H(Z)$, which differs from Shannon's entropy of $X$ as regards the number of categories. When the order is not preserved, this measure does not depend on the spatial configuration of co-occurrences. Therefore, in this case, $Z$ maintains the information of $X$ and the corresponding entropies are strictly related (see Section \ref{sec:sim_resshan} for details). Conversely, when the order is preserved, the entropy of $Z$ depends not only on the pmf of $X$, but also on the spatial order of its realizations.

All contributions that introduce space in entropy measures based on $Z$ make use of a definition of neighbourhood which, even in the simplest case of sharing a border, needs the construction of an adjacency matrix $A$, which, for a generic degree $m$, generalizes to a hypercube in the $m$-dimensional space. The definition of $A$ implies that the univariate distribution used in entropies are conditional, i.e. $p_{Z|A}=(p(z_1|A), \dots, p(z_{R_m}|A))'$. Realizations of $Z|A$ form a subset of realizations of $Z$ that only includes co-occurrences identified by non-zero elements of $A$, i.e. conditioning on a fixed neighbourhood. 

In most works on regular lattice data (see, for instance, \citeauthor{oneill}, \citeyear{oneill}, \citeauthor{contagion}, \citeyear{contagion}, \citeauthor{riitters} \citeyear{riitters}), co-occurrences are defined as ordered couples of contiguous realizations of $X$, where the term "contiguous" in this case means "sharing a border". Thus, $m=2$ and a contiguity matrix is built, here denoted by $O$; consequently, the variable of interest is $Z|O$ with $R^{o}_2=I^2$ categories.

\citet{oneill} propose one of the early spatial entropy indices, computing a Shannon's entropy (\ref{eq:shann}) for the variable $Z|O$
\begin{equation}
H(Z|O)=E\left[I\left(p_{Z|O}\right)\right]=\sum_{r=1}^{R^o_2} p(z_{r}|O)\log\left(\frac{1}{p(z_{r}|O)}\right).
\label{eq:oneill}
\end{equation}
The entropy ranges from 0 to $\log(R^o_2)$; the index maximum is reached when the pmf $p_{Z|O}$ is uniform.

Other measures based on the construction of $Z|O$ start from the concept of contagion, the conceptual opposite to entropy. The Relative Contagion index $RC$ \citep{contagion} was proposed as
\begin{equation}
RC(Z|O)=1-H_{norm}(Z|O)=1-\frac{1}{\log(R^o_2)}\sum_{r=1}^{R^o_2} p(z_{r}|O)\log\left(\frac{1}{p(z_{r}|O)}\right)
\label{eq:rc}
\end{equation}
where the idea of normalized measure comes from (\ref{eq:relH}): the second term is the normalized entropy of $Z|O$, via the multiplication of (\ref{eq:oneill}) by the appropriate $B=1/\log(R^o_2)$. Then its complement to 1 is computed in order to measure relative contagion: the higher the contagion between categories of $Z|O$, the lower the entropy.

\cite{riitters} derive RC indices with or without preserving the order of elements in the couple. In the latter case, the number of categories for $m=2$ is $R^{no}_2=(I^2+I)/2$, a special case of the binomial coefficient $R^{no}_m=\binom{I+m-1}{m}$, and the normalization constant changes accordingly into $B=1/\left(\log(I^2+I)-\log(2)\right)$. 

A negative characteristic of the RC index, as noted by \cite{parresol} is that, as all normalized measures, it is not able to distinguish among contexts with different numbers of categories. While a dataset with $I=2$ has a lower entropy than a dataset with $I=100$, a normalized index does not account for that. For this reason, \citeauthor{parresol} suggest to go back toward an unnormalized version of (\ref{eq:rc}):
\begin{equation}
C(Z)=-H(Z|O)=\sum_{r=1}^{R^o_2} p(z_{r}|O)\log(p(z_{r}|O))
\label{eq:gamma}
\end{equation}
thus ranging from $-\log(R^o_2)$ to $0$.

The above measures are inspired by very different conceptualizations but are computed as linear transformations of the common starting quantity (\ref{eq:oneill}).

\cite{leibovici09} and \cite{leibovici14} propose a richer measure of entropy by extending $H(Z|O)$ in two ways. Firstly, $Z$ can now represent not only couples, but also triples and further degrees $m$ of co-occurrences. The authors only develop the case of ordered co-occurrences, so that the number of categories of $Z$ is $R^o_m=I^m$. Secondly, space is now allowed to be continuous, so that areal as well as point data might be considered and associations may not coincide with contiguity; therefore the concept of distance between occurrences replaces the concept of contiguity between lattice cells. A distance $d$ is fixed, then co-occurrences are defined for each $m$ and $d$ as $m$-th degree simultaneous realizations of $X$ at any distance $d^{*} \le d$, i.e. distances are considered according to a cumulative perspective; this way an adjacency hypercube $L_d$ is built and the variable of interest is $Z|L_d$. Then, Leibovici's spatial entropy is
\begin{equation}
H(Z|L_d)=E\left[I\left(p_{Z|L_d}\right)\right]=\sum_{r=1}^{R^o_m} p(z_r|L_d)\log\left(\frac{1}{p(z_r|L_d)}\right).
\label{eq:leib}
\end{equation}
The probability $p(z_r|L_d)$ is again the discrete element of a univariate pmf $p_{Z|L_d}$, i.e. computed for a distribution conditional on $L_d$. Entropy (\ref{eq:leib}) can be normalized using $B=1/\log(R^o_m)$. In the case of lattice data, O'Neill's entropy (\ref{eq:oneill}) is obtained as a special case when $m=2$ and $d$ equals the cell's width.

\subsection{Overall comments}

The categorical variable $X$ of Section 2 is not used in entropies (\ref{eq:spaten}) and (\ref{eq:karl}), as the information on the different categories is lost. The phenomenon of interest $F$ may coincide with a specific category of $X$, say $F=X_i^*$, and $H_{KC}(X_i^*)$ may be computed to assess the spatial configuration of the realizations of $X_i^*$. Thus, for a categorical $X$, $I$ different $H_{KC}(X_i^*)$ can be computed, but there is no way to synthesize them into a single spatial entropy measure for $X$. A similar approach exploiting neighbourhood in terms of spatial proximity is also used in several sociological works to properly define an entropy-based segregation index among population groups able to include spatial information of locations (see, e.g., \citeauthor{reardon}, \citeyear{reardon}).

With respect to the measures proposed in Section \ref{sec:spatent}, the advantage of an approach based on the construction of $Z$ is to maintain the information about all categories of $X$. It holds however two main limits: firstly, entropies are not decomposable, so there are no partial terms to investigate; secondly, they are based on univariate distributions, so all the interesting properties related to bivariate distributions cannot be enjoyed. Two substantial differences  can be appreciated when using an appropriate adjacency matrix to build the realizations of $Z|A$ with respect to \citeauthor{karlstrom}'s approach of Section \ref{sec:kc}. First of all, in \citeauthor{karlstrom}'s approach $p_g$ does not depend on $A$, which is needed in the further step to include the neighbouring values, i.e. to derive $\widetilde{p}_g$; on the contrary, in the approach proposed in Section \ref{sec:Z}, $A$ is needed from the beginning to switch from $X$ to $Z|A$ and to define the proper pmf $p_{Z|A}$ itself. Secondly, since $p_g$ takes values over a location $g$, the other probabilities $p_{g'}$, $g'\ne g$ are used to compute $\widetilde{p}_g$ in the neighbourhood of each $g$. Conversely, in the approach based on the construction of $Z|A$, probabilities $p(z_r|A)$ are not referred to a specific location.

\section{Additive spatial entropy measures exploiting bivariate properties}
\label{sec:nostra}

All the previously listed indices are challenging attempts to include space into entropy measures; nevertheless, some open questions remain. An important limitation is that each index is computed for only one adjacency matrix, i.e. by fixing the neighbourhood of interest in advance. This is linked to the fact that all entropies of Section \ref{sec:reviewspace} are based on univariate distributions and cannot take advantage of the bivariate properties presented in Section \ref{sec:prop}. The use of bivariate distributions would allow the property of additivity to be exploited for a global index by using a rigorous probabilistic approach. Moreover, there is a need to build spatial entropy measures exploiting the relationship of the study variable with space. Under this perspective, proper spatial entropy measures are expected to:
\begin{enumerate}
\item[a)] maintain the information about the categories of $X$, e.g. by exploiting the trasformed variable $Z$ as in Section \ref{sec:Z},
\item[b)] consider different distance ranges simultaneously, by including an additional study variable representing space to enjoy the properties of bivariate entropy measures,
\item[c)] quantify the overall role of space,
\item[d)] be additive and decomposable, i.e. satisfy partial and global properties as in Section \ref{sec:kc}.
\end{enumerate}

All the above points are accomplished using residual entropy (\ref{eq:residYX}) and its relationship with Shannon's entropy (\ref{eq:shann}); residual entropy is the proper quantity able to summarize partial entropy measures, conditional on a specific value of the second variable, into a global one. Partial entropies, conditional expectations themselves, are weighted by their probabilities, helping to appreciate the relevance of uncertainty and to switch from explorative analysis to statistical inference. The same properties a) to d) are enjoyed by the quantity known as mutual information (\ref{eq:mutual}), which receives a novel interpretation when space is taken into account. 

The realizations of $X$ are assumed to occur over a discretized area, say a grid (though the following is applicable to non-regular lattices and point data as well), and distances between areas are represented by Euclidean distances between centroids. Co-occurrences of $X$ are used to build $Z$, and the degree of such co-occurrences (couples, triples, or further degrees, i.e. $m=1,\ldots ,M$) is fixed exogenously, driven by the researcher's experience. Different structures have different merits, discussing them is beyond the purpose of this work and conclusions are independent of such choice. The categories of the transformed variable $Z$ derive from unordered associations: ordering occurrences does not appear sensible, since spatial neighbourhoods do not generally have a direction. Moreover, neglecting the order of occurrences ensures a one-to-one correspondence between $H(X)$ and $H(Z)$. Conversely, if order is preserved, different $H(Z)$ can be obtained for different spatial configurations of the same series of realizations of $X$, as discussed in Section \ref{sec:reviewspace}. This encourages the choice of considering unordered occurrences as the most appropriate: in the case of $m=2$, a spatial measure should consider the couple $z_{r}=(x_i,x_j)$ equal to the couple $z_{r}=(x_j,x_i)$, and analogously for further degrees of co-occurrences, and this is what will be done in the remainder of this Section.

As above mentioned, a novelty of the proposed measures lies in the introduction of a second discrete variable $W$, that represents space by classifying the distances at which co-occurrences take place. These exogenous classes $w_k$, with $k=1,\dots,K$, cover all possible distances within the observation window and have a pmf $p_W=(p(w_1), \dots, p(w_K))'$; $p(w_k)$ is the probability associated to the $k$-th distance range. Once the degree $m$ of co-occurrences is fixed, each distance category $w_k$ implies the choice of a different adjacency matrix $A_k$ (a hypercube in the $m$-dimensional space for $m>2$) for the associations of $X$ that define $Z|A_k$. Therefore, $p_{Z|A_k}$ may equivalently be written as the $R^{no}_m \times 1$ vector $p_{Z|w_k}$, and the set of $K$ conditional distributions can be collected in a $R^{no}_m \times K$ matrix
\begin{equation}
p_{Z|W}=\begin{bmatrix} p_{Z|w_1} & p_{Z|w_2} & \cdots & p_{Z|w_K} \end{bmatrix}.
\end{equation}

Consequently, the discrete joint pmf $p_{ZW}$ can be represented by a $R^{no}_m \times K$ matrix:
\begin{equation}
p_{ZW}=p_{Z|W}diag\left(p_W\right).
\end{equation}
This decomposition is relevant to stress the logical relationship between $Z$ and $W$: $W$ influences $Z$ and not viceversa. It confirms that the marginal pmf of $W$ and the set of distributions of $Z$ conditional on the values of $W$ are the proper quantities to obtain entropy measures exploiting properties of bivariate distributions. 

\subsection{An innovative view warranting additivity: spatial residual entropy}

Under the setting defined at the beginning of this Section, entropy has to be computed on the $Z|w_k$; the number of categories is $R^{no}_m$, and the strength of spatial influence is determined by  the $K$ values of $W$. These elements permit to write the entropy measure called spatial global residual entropy $H(Z)_W$, which reinterprets (\ref{eq:residYX}) as follows
\begin{equation}
\begin{split}
H(Z)_W=E[H(Z|w_k)]=E[E\left(I\left(p_{Z|w_k}\right)\right)]&=\sum_{k=1}^{K}p(w_k)\sum_{r=1}^{R^{no}_m}p(z_r|w_k)\log{\left(\frac{1}{p(z_r|w_k)}\right)}\\
&=\sum_{r=1}^{R^{no}_m}p(z_r,w_k)\log{\left(\frac{1}{p(z_r|w_k)}\right)}.
\end{split}
\label{eq:residZW}
\end{equation}
The components of (\ref{eq:residZW})
\begin{equation}
H(Z|w_k)=E[I\left(p_{Z|w_k}\right)]=\sum_{r=1}^{R^{no}_m}p(z_r|w_k)\log{\left(\frac{1}{p(z_r|w_k)}\right)}
\label{eq:residZW_loc}
\end{equation}
have a crucial meaning and, from now on, are named spatial partial residual entropies, where "partial" corresponds to a specific distance class $w_k$. They are computed starting from the conditional pmf $p_{Z|w_k}$. When these measures are multiplied by the probability $p(w_k)$, they allow spatial global residual entropy (\ref{eq:residZW}) to enjoy the additive property, as (\ref{eq:residZW}) can be written as
\begin{equation}
H(Z)_W=\sum_{k=1}^{K}p(w_k)H(Z|w_k).
\label{eq:nostra_add}
\end{equation}

The additive relationship (\ref{eq:nostra_add}) holds, as the spatial global residual entropy (\ref{eq:residZW}) is obtained by weighting the spatial partial residual entropies with the probabilities of the conditioning variable $W$, and is relevant: the spatial global residual entropy tells how much information is still brought by $Z$ after removing the effect of the spatial configuration $W$. Partial entropies show how distances contribute to the entropy of $Z$.


The main innovation of the proposed spatial residual entropy perspective is that it allows entropy measures illustrated in Section \ref{sec:Z} to be generalized through the formulation of each different spatial partial residual entropy (\ref{eq:residZW_loc}). Indeed, fixing a degree $m=2$ and a distance class $w_k$ implies a definition of an adjacency matrix $A_k$ which can be the contiguity $O$ as in (\ref{eq:oneill}) when $w_k=[0,1]$, or a $L_d$ as in (\ref{eq:leib}) based on a distance range $w_k=[0,d]$.

\subsection{Deepening the concept of mutual information}

An immediate consequence of relying on residual entropy is the possibility to isolate the mutual information of $Z$ and $W$, from now on named spatial mutual information, according to (\ref{mutrul}) by subtracting the spatial global residual entropy $H(Z)_W$ from  $H(Z)$, Shannon's entropy of $Z$:
\begin{equation}
MI(Z, W)=H(Z)-H(Z)_W.
\label{eq:residZW_mut}
\end{equation}
Shannon's entropy of $Z$ is computed by using the univariate marginal $p_Z$, that does not depend on any adjacency matrix. 

Spatial mutual information is defined, similarly to (\ref{eq:mutual}), as
\begin{equation}
MI(Z,W)=E\left[I\left(\frac{p_Zp_W}{p_{ZW}}\right)\right] =\sum_{r=1}^{R^{no}_m}\sum_{k=1}^{K}p(z_r,w_k)\log{\left(\frac{p(z_r,w_k)}{p(z_r)p(w_k)}\right)}.
\label{eq:spatial_mutual}
\end{equation}
It is a Kullback-Leibler distance $D_{KL}(p_{ZW}||p_Zp_W)$ and the component of $H(Z)$ due to the spatial configuration $W$. Spatial mutual information may be additively decomposed the same way as spatial global residual entropy (\ref{eq:residZW}), so that the contribution of space can be quantified at every distance range $w_k$:
\begin{equation}
MI(Z,W)=\sum_{k=1}^{K} p(w_k)\sum_{r=1}^{R^{no}_m}p(z_r|w_k)\log{\left(\frac{p(z_r|w_k)}{p(z_r)}\right)},
\label{eq:partial_mutual}
\end{equation}
where the $k$-th partial term, analogously to (\ref{eq:residZW_loc}), is now named spatial partial information
\begin{equation}
PI(Z,w_k)=E\left[I \left(\frac{p_Z}{p_{Z|w_k}}\right) \right]=\sum_{r=1}^{R^{no}_m}p(z_r|w_k)\log{\left(\frac{p(z_r|w_k)}{p(z_r)}\right)}.
\label{eq:partialterm_mut}
\end{equation}

Each partial term is a Kullback-Leibler distance $D_{KL}(p_{Z|w_k}||p_Z)$ that quantifies the contribution to the departure from independence of each conditional distribution $p_{Z|w_k}$. In the special case of independence between $Z$ and a distance class $w_k$, $p(z_r|w_k)=p(z_r)$ and the contribution of the corresponding partial term to the spatial mutual information is null. The additive relationship is respected the same way as for the spatial residual entropy, once the $PI$s are weighted by the probabilities $p(w_k)$:
\begin{equation}
MI(Z,W)=\sum_{k=1}^{K}p(w_k)PI(Z,w_k).
\label{eq:mut_sumPI}
\end{equation}
Again, (\ref{eq:partial_mutual}) is expressed in terms of $w_k$, due to the logical order between $w_k$ and $z_r$. 

\subsection{Advances in interpreting spatial entropy measures}
\label{sec:nostra_advances}

Expression (\ref{eq:residZW_mut}) now takes a new substantial meaning: the entropy of $Z$, $H(Z)$, may be decomposed into spatial mutual information, quantifying the role of space, and spatial global residual entropy, quantifying the remaining information brought by $Z$:
\begin{equation}
H(Z)=MI(Z,W)+H(Z)_{W}.
\end{equation}
The more $Z$ depends on $W$, i.e. the more the realizations of $X$ are (positively or negatively) spatially associated, the higher the spatial mutual information. Conversely, when the spatial association among the realizations of $X$ is weak, the entropy of $Z$ is mainly due to the spatial global residual entropy.

For the sake of interpretation and diffusion of the results, a ratio can be built that allows to quantify the role of space in proportional terms. The quantity
\begin{equation}
MI_{prop}(Z,W)=\frac{MI(Z,W)}{H(Z)}=1-\frac{H(Z)_W}{H(Z)}
\label{eq:mutprop}
\end{equation} 
ranges in $[0,1]$ and is able to quantify the contribution of space in the entropy of $Z$ as a proportion of the marginal entropy. If, e.g., $MI_{prop}(Z,W)=0.6$, it can be concluded that 60\% of the entropy of $Z$ is due to the specific spatial configuration. Similarly, $H(Z)_W/H(Z)$ gives the proportion of the entropy of $Z$ due to sources of heterogeneity other than space. This highlights that both $MI(Z,W)$ and $H(Z)_W$ can potentially vary in the whole range of $H(Z)$, but the value taken by $H(Z)$ constitutes an upper limit.

Moreover, the terms in $H(Z)$ can be further decomposed, exploiting (\ref{eq:nostra_add}) and (\ref{eq:mut_sumPI}), as
\begin{equation}
H(Z)=\sum_{k=1}^{K}p(w_k)\left[PI(Z,w_k)+H(Z|w_k)\right],
\end{equation}
where the contribution of each term in explaining the relationship between $Z$ and $W$ is isolated. This way, Shannon's entropy $H(Z)$ can be written in additive form by exploiting the bivariate properties of entropy.

\section{A comparative study of spatial entropy measures}
\label{sec:sim}

Section \ref{sec:nostra} has already presented the theoretical properties of the proposed measures, which state their superiority as spatial entropy indices. Unlike traditional measures, all referred to a univariate approach and some based on a definition of a single adjacency matrix (hypercube), spatial residual entropy (\ref{eq:residZW}) and spatial mutual information (\ref{eq:spatial_mutual}) consider different matrices (hypercubes) to cover all possible distances and exploit the bivariate properties of entropy to summarize what is known. It has been shown that all measures in Section \ref{sec:Z} can be derived as special cases of one spatial partial residual entropy (\ref{eq:residZW_loc}). The spatial entropy indices discussed in Section \ref{sec:reviewspace} and \ref{sec:nostra} need to be further investigated in order to identify their main properties and the different contexts of application.
 The behaviour of the proposed entropy indices compared to the other measures is assessed in what follows, and their flexibility and informativity are investigated. Therefore, in this comparative analysis, several datasets are generated under different scenarios to compute Batty's entropy (\ref{eq:spaten}) and \citeauthor{karlstrom}'s entropy (\ref{eq:karl}). Then, O'Neill's entropy (\ref{eq:oneill}) and its generalization Leibovici's (or co-occurrence) entropy (\ref{eq:leib}) are also assessed. Finally, the spatial global residual entropy (\ref{eq:residZW}) and the spatial mutual information (\ref{eq:spatial_mutual}) are computed, as well as their partial components. The thorough comparison of this wide set of entropy measures shows that spatial residual entropy overcomes the traditional measures as regards completeness, since it succeeds in synthesizing many relevant features. It is also the main tool for computing spatial mutual information, able to point out the overall role of space.

For simplicity of presentation, discrete space and a regular grid are considered. It is to point out that space can be discretized as wished, as long as a distance measure between areas is suitably defined. Additionally, space is allowed to be continuous and areas may be replaced by points; in this case, $W$ would represent the distance between points themselves, and all entropy measures would be defined accordingly. When dealing with the transformed variable $Z$, $m=2$ is assumed and the number of categories is simply named $R^{o}$ or $R^{no}$ according to order preservation.

Sections \ref{sec:simdesign} and \ref{sec:cumput} illustrate the design of the study: firstly, the data generation procedure is introduced, then, estimation of the necessary probability distributions is presented. Results are shown for the non-spatial Shannon's entropy in Section \ref{sec:sim_resshan}; afterwards, results for all entropy measures on data with two categories are summarized in Section \ref{sec:sim_resX2}. The main results for extensions to data with more than two categories are in Section \ref{sec:sim_resX5X20}.

\subsection{Data generation}
\label{sec:simdesign}

\begin{figure}
\centering \includegraphics[width=.75\textwidth]{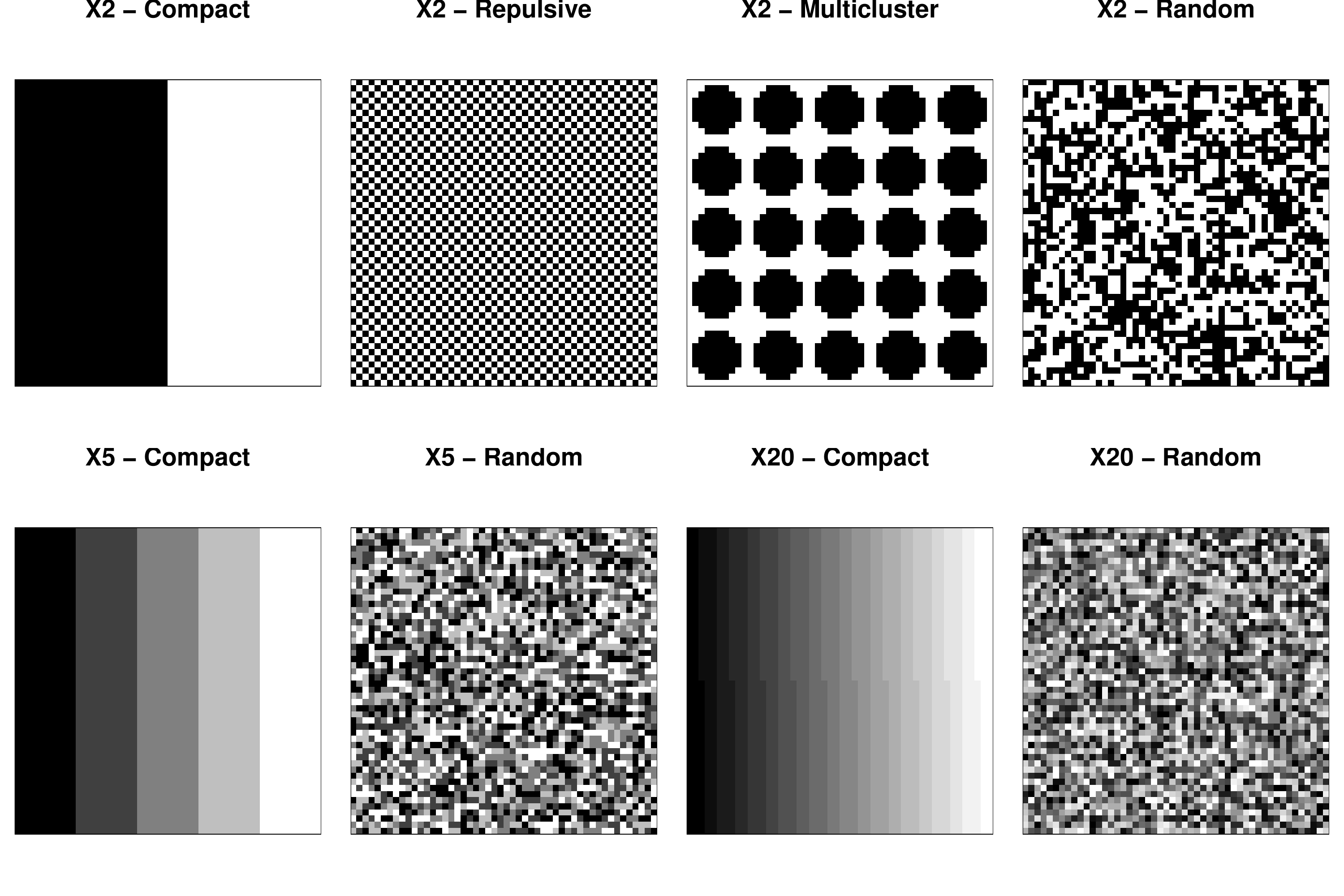}
          \caption{Data generated for all simulation scenarios under a uniform distribution of $X$.}\label{fig:1}
\end{figure}

Let us consider $N=2500$ realizations of a categorical variable $X$ by randomly setting the pmf $p_X$ and then generating values from a multinomial distribution $Mn(N,p_X)$. In accordance to the choice of using a regular grid, the realizations are arranged in $50\times 50$ pixels over a square window. Without loss of generality, each pixel is assumed to be a $1\times 1$ square, therefore the observation window is $50 \times 50$ units. Three options for the number of categories $I$ are covered: 2 categories ($X2$), 5 categories ($X5$) and 20 categories ($X20$). Categories are coded with integers from 1 to $I$ and, when needed, represented by different grey intensities going from black to white. The simulated sequence of 2500 values is organized according to different spatial configurations, as they are expected to produce different entropy values.
For $X2$, four different scenarios are built:
\begin{enumerate}
\item compact - a spatially strongly clustered distribution
\item repulsive - a spatially regular distribution, tending to a chessboard configuration
\item multicluster - a spatial configuration presenting 25 small clusters of about the same size
\item random - a pattern with no spatial correlation whatsoever.
\end{enumerate}
As regards $X5$ and $X20$, two scenarios are considered, which represent the two extreme entropy situations: the compact and the random ones. Indeed, when many unordered categories are present over a window, a repulsive pattern is uninteresting, as it would be very hard to distinguish it from a random one. For a similar reason, a multicluster configuration is not built. Hence, eight simulated scenarios are investigated, each replicated $1000$ times.
A dataset generated under the hypothesis of uniform distribution $U_X$ among the categories is also built as a special case for each of the eight scenarios, as the 1001$-th$ simulation with $p(x_i)=1/I$ for every $i$; it is displayed in Figure \ref{fig:1}. For the multicluster configuration, the 25 cluster centroids are, in this special case, also forced to be uniformly distributed over the square window.

\subsection{Computation of entropy components}
\label{sec:cumput}

Generated data are used to compute quantities for the entropy measures to be assessed and, in particular, probabilities are estimated as proportions of observed data, as follows.

When restricting to a regular grid, i.e. the ultimate partition of the observation window, each unit contains one (and one only) realization of $X$ and the pixel size is 1. Let the generic pixel be labelled by $u$, $u=1, \dots, N$ and let $x_u$ denote the value of $X$ in pixel $u$. For Shannon's entropy (\ref{eq:shann}) the probabilities $p(x_i)$ for each category are estimated by the proportion of pixels where $x_i$ is observed: 
\begin{equation}
\widehat{p}(x_i)= \frac{\sum_{u=1}^N\mathbf{1}(x_u=x_i)}{N}.
\end{equation}

\begin{figure}
\centering
    \includegraphics[width=.75\textwidth]{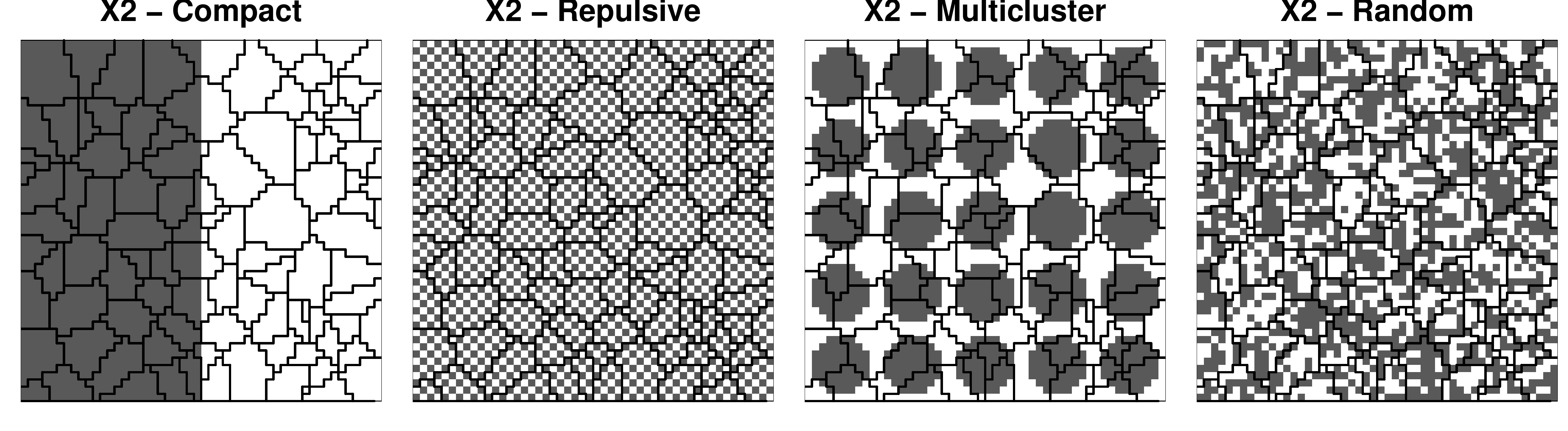}
          \caption{Four scenarios under the uniform distribution for $X2$, with partition in 100 areas.}\label{fig:batty_unif_part}
\end{figure}

Batty's and \citeauthor{karlstrom}'s entropies cannot be computed directly on the pixel grid, since only one realization of $X$ occurs over each pixel. The studied phenomenon is here defined as the occurrence of 1-valued pixels (black pixels in figures) over a fixed area, i.e. $F=X_1$; indices are computed on data generated for $X2$.
The window is partitioned in $G=100$ fixed areas of different size, where the size of each area, $T_g$, $g=1, \dots, 100$, is the number of contained pixels. The superimposition of the areas over the data matrices is shown in Figure \ref{fig:batty_unif_part} for the datasets generated under a uniform distribution of $X2$ for the different spatial configurations shown in Figure \ref{fig:1}. The probabilities $p_g$ are estimated in each of the $1000$ simulations as the proportion of 1-valued pixels over the areas:
\begin{equation}
\widehat{p}_g=\frac{c_g}{C}
\end{equation} 
where $c_g$ is the number of 1-valued pixels in area $g$ and $C$ is the total number of 1-valued pixels, so that $\sum_g \widehat{p}_g=1$.
For computing \citeauthor{karlstrom}'s entropies, 4 neighbourhood distances are considered between the areas' centroids to quantify $I(\widetilde{p}_g)$: $d_0=0$ (no neighbours other than the pixel itself), $d_1=2$, $d_2=5$ and $d_3=10$. The definition of the adjacency matrix $A$ based on the fixed distance $d$ and the estimated quantities $\widehat{p}_g$ allow to compute the estimates of $\widetilde{p}_g$ as averages of the neighbouring estimated probabilities
\begin{equation}
\widehat{\widetilde{p}}_g=\sum_{g'=1}^G a_{gg'}\widehat{p}_{g'}=\frac{\sum_{g' \in \mathcal{N}(g)}\widehat{p}_{g'}}{|\mathcal{N}(g)|}.
\end{equation} 

For the measures based on the transformation $Z$ of the study variable $X$, the variable $Z$ has a different number $R$ of categories according to the number of categories of $X$ and whether the order is preserved; options for $I$ and $R$ values and the corresponding entropy maxima (i.e. the entropy obtained under a uniform distribution for $X$ and $Z$ respectively) are reported in Table \ref{tab:Zcat}. When the order is not preserved, the entropy range is smaller.
\begin{table}
\caption{\label{tab:Zcat}Categories and entropy maxima for $X$ and $Z$.}
\centering
\fbox{
\begin{tabular}{c|c|c||c|c|c}
    \multicolumn{3}{c||}{No of categories} & \multicolumn{3}{c|} {Entropy maxima} \\
    \hline
    $X$ & \multicolumn{2}{|c||}{$Z$} & $X$ & \multicolumn{2}{|c|}{$Z$} \\
    \hline
   $I$ & $R^o$ & $R^{no}$ & $\log(I)$ & $\log(R^o)$ & $\log(R^{no})$ \\
    \cline{1-6}
    2 & 4 & 3 & 0.69 & 1.38 & 1.10 \\
    5 & 25 & 15 & 1.61 & 3.22 & 2.71 \\
    20 & 400 & 210 & 2.99 & 5.99 & 5.35 \\
\end{tabular}}
\end{table}

Co-occurrences are built according to the specific adjacency matrix employed in each case, as described in Section \ref{sec:Z}. The cell centroids are used to measure distance between pixels, consequently the distance between contiguous pixels is $1$ and the distance to farther cells along the cardinal directions belongs to the set of integers $\mathbb{Z}^+$.
The adjacency matrix $A_k$ is $N \times N$ and $|\mathcal{N}(u)_k|=\sum_{u'=1}^N a_{uu',k}$ is the number of $Z$ observations built using the neighbourhood $\mathcal{N}(u)_k$ of pixel $u$. 
The rule of moving rightward and downward is adopted along the observation window in order to identify adjacent couples. 

A general method for estimating $p_{Z|w_k}$ is proposed, which can be applied, after choosing a suitable adjacency matrix, to all measures in Section \ref{sec:Z}. 
Let $Q_k$ denote the number of observed couples over the dataset for each category $w_k$, corresponding to the sum of all unit values over the matrix $A_k$: $Q_k=\sum_{u=1}^N |\mathcal{N}(u)_k|=\sum_{u=1}^N \sum_{u'=1}^N a_{uu',k}$. All observed $Z|w_k$ over the dataset are arranged in the rows of a $Q_k \times 2$ matrix $Z^{obs}_k=[Z^{(1)}_k, Z^{(2)}_k]$. The first column of $Z^{obs}_k$ is obtained by taking each pixel value and replicating it as many times as the cardinality of its neighbourhood:
\begin{equation}
Z^{(1)}_k=\begin{bmatrix} x_1 \cdot 1_{|\mathcal{N}(1)_k|}\\ x_2 \cdot 1_{|\mathcal{N}(2)_k|}\\\vdots \\ x_N \cdot 1_{|\mathcal{N}(N)_k|} \end{bmatrix}
\label{eq:Zobs_1}
\end{equation}
where each $1_{|\mathcal{N}(u)_k|}$ is a $|\mathcal{N}(u)_k|$-dimensional vector of 1s. The second column $Z^{(2)}_k$ is built selecting, for each pixel, the neighbouring values via $A_k$. Let us define the $N \times N$ selection matrix $\widetilde{A}_k$, substituting zeros in $A_k$ with missing values. Let us also define $vec(X)$ as a $N \times 1$ vector stacking all realizations of $X$. An element-wise product is run between $vec(X)$ and the $u$-th row of $\widetilde{A}_k$, denoted by $vec(X) \cdot  \widetilde{A}_{u., k}$. Thus, a $|\mathcal{N}(u)_k|$-dimensional vector is returned, containing the values of the pixels neighbouring $u$:
\begin{equation}
Z^{(2)}_k=\begin{bmatrix} vec(X) \cdot \widetilde{A}_{1., k}\\
vec(X) \cdot \widetilde{A}_{2., k}\\
\vdots \\
vec(X) \cdot \widetilde{A}_{N., k} \end{bmatrix}.
\label{eq:Zobs_2}
\end{equation}
The $Q_k$ realizations of $Z^{obs}_k$ present at most $R^{no}$ categories, indexed by $r=1, \dots, R^{no}$. Their relative frequencies are used to compute $\widehat{p}(z_r|w_k)$. The marginal $p_Z$ may be estimated by marginalizing out $W$, or by building a special adjacency matrix $A$ that takes value 1 everywhere except for the main diagonal (such a matrix is indeed the sum of all $A_k$, $k=1, \dots, K$). The estimated pmf $\widehat{p}_Z$ is used to compute Shannon's entropy $H(Z)$.

For the computation of O'Neill's spatial entropy (\ref{eq:oneill}), the above method employs the adjacency matrix $O$; for Leibovici's entropy (\ref{eq:leib}), the adjacency matrix $L_d$ is used and results are shown for $d=2$. When computing the spatial residual entropy (\ref{eq:residZW}), the variable $W$ is built with fixed categories $w_k$: $w_1=[0,1]$, $w_2=]1,2]$, $w_3=]2,5]$, $w_4=]5,10]$, $w_5=]10,20]$, $w_6=]20,30]$ and $w_7=]30, 50\sqrt{2}]$ (where $50\sqrt{2}$ is the diagonal length, i.e. the maximum distance over a square of side 50), covering all possible distances for couples over the dataset. For each distance $w_k$, a specific adjacency matrix $A_k$ is built. Therefore, $K$ different $A_k$ and $Z^{obs}_k$ are built using (\ref{eq:Zobs_1}) and (\ref{eq:Zobs_2}), and $K$ different conditional distributions $\widehat{p}_{Z|w_k}$ are obtained. Finally, an estimate for $p_W$ is needed: for each $k$, $\widehat{p}(w_k)=Q_k/Q$ represents the proportion of couples within distance range $w_k$ with respect to the total number of couples $Q=\sum_k Q_k$. A summary of the characteristics of entropy measures based on $Z$ is shown in Table \ref{tab:Zcat2}, highlighting the specific adjacency matrix for each index.

\begin{table}
\caption{\label{tab:Zcat2}Peculiarities of entropy measures based on $Z$.}
\centering
\fbox{
\begin{tabular}{c|c|c|c}
    Entropy & No cat. $Z$ & Adjacency matrix & No obs. couples  \\
    \hline
$H(Z|O)$ &  $R^o$& $O$ (contiguity)  & $Q_O=4900$ \\
$H(Z|L_d)$ &  $R^o$ & $L_2$ (up to distance $2$) & $Q_{L_2}=14502$ \\
$H(Z|w_k)$ and $PI(Z,w_k)$ &  $R^{no}$& $A_1$ (at distance $w_1$) & $Q_1=4900$  \\
 &   &$\vdots$ & $\vdots$\\
  & & $A_7$ (at distance $w_7$) & $Q_7=1191196$ \\
\end{tabular}}
\end{table}

All the above mentioned indices are computed for each scenario over the 1000 generated datasets, plus the special case of uniform distribution among the $X$ categories.
In the presentation of the results, boxplots are employed to summarize the distribution of a specific index; stars highlight results achieved under the uniform distribution of $X$, while the dashed lines mark the indices' maxima.

\subsection{Results for Shannon's entropy}
\label{sec:sim_resshan}

\begin{figure}
\centering
     \includegraphics[width=.75\textwidth]{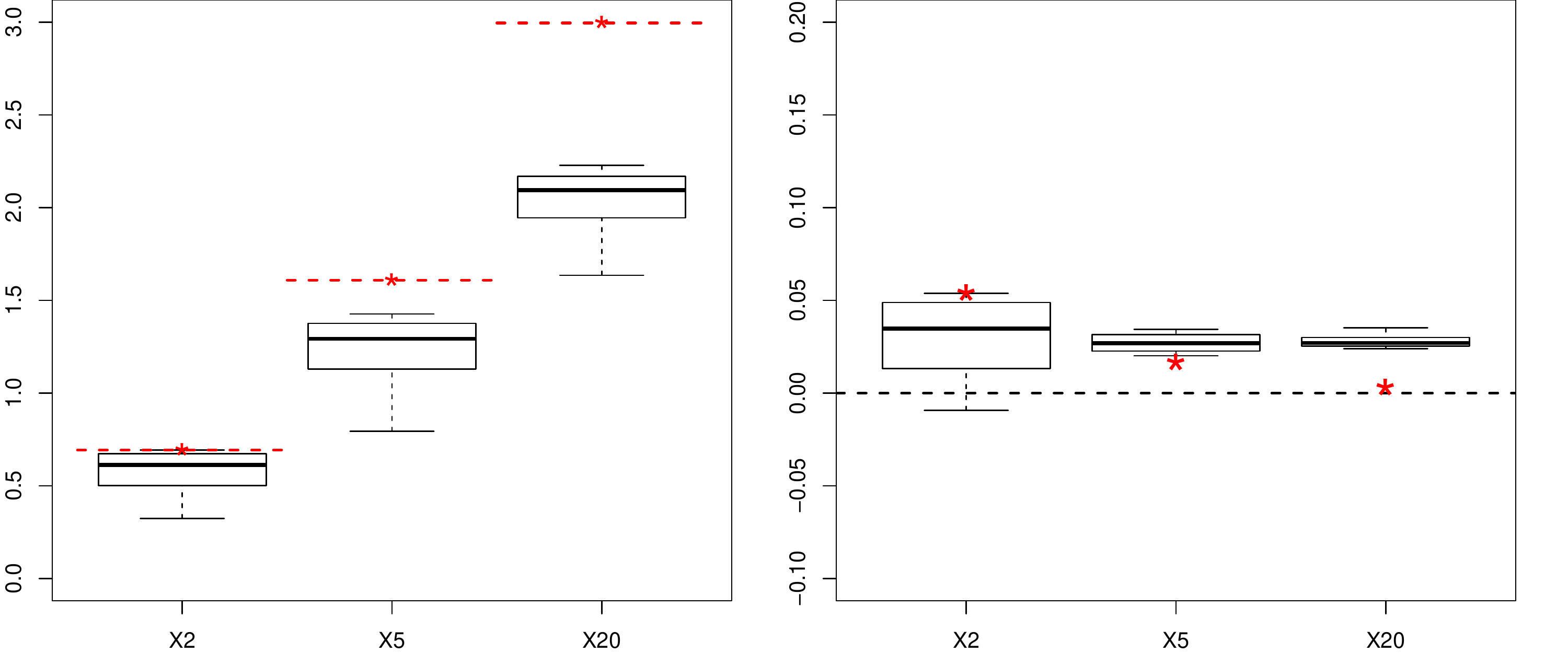}

     \caption{Left panel: Shannon's entropy, 1000 simulations; each dashed line corresponds to the index maximum. Right panel: difference between normalized Shannon's entropy of $X$ and $Z$. Each star identifies the entropy value computed on a uniformly distributed $X$. }\label{fig:sh}
\end{figure}

Figure \ref{fig:sh}, left panel, shows Shannon's entropy (\ref{eq:shann}) for $X2$, $X5$ and $X20$.
Since Shannon's entropy is non-spatial, it only depends on the generated outcomes and not on their spatial distribution, therefore there are no distinct measures for the different spatial configurations proposed: entropy only depends on $X$ categories' proportions and not on space. Entropy is higher as the number of categories increases: the greater diversity (i.e. number of categories), the higher entropy. Moreover, the empirical probability intervals do not overlap, therefore the index is effective in distinguishing among contexts with different numbers of categories. Should the normalized version of the index be computed, interpretation would be easier, but it would be impossible to distinguish among $X2$, $X5$ and $X20$. In all cases, the unnormalized entropy maximum is $\log(I)$ and is reached when the distribution of $X$ is uniform.
Shannon's entropy can be analogously computed on $Z$ without order preservation, so that it does not depend on any adjacency matrix as it is non-spatial. It is computed via the $\widehat{p}_Z$ of Section \ref{sec:cumput} and compared to Shannon's entropy of $X$ in Figure \ref{fig:sh}, right panel. The Figure highlights that $Z$ brings the same information as $X$, since the two normalized entropy measures tend to have both the same behaviour and the same range of values; the difference is very small and becomes negligible as $I$ increases.
Expression $H(Z)$ is used in Section \ref{sec:sim_resX2} for computing the mutual information.

\subsection{Results for binary data}
\label{sec:sim_resX2}

This Section thoroughly illustrates the performance of the measures of Section \ref{sec:reviewspace} when applied to binary data. 
According to the data generation description in Section \ref{sec:simdesign}, the variable $X$ assumes values $x_1=1$ (black pixels) or $x_2=2$ (white pixels) and values are drawn from a binomial distribution as a special case of the multinomial distribution. The variable $Z$ with order preservation has $R^o=4$ categories $z_1=(1,1)$, $z_2=(2,2)$, $z_3=(1,2)$ and $z_4=(2,1)$; when order is not preserved, $R^{no}=3$ as the last two categories are undistinguishable.

Section \ref{sec:simres_batty} relates to the measures of Section \ref{sec:spatent}, and Section \ref{sec:simres_Z} to those of Section \ref{sec:Z}; in addition, Section \ref{sec:simres_resid} refers to the proposals of Section \ref{sec:nostra}.

\subsubsection{Batty's and \citeauthor{karlstrom}'s entropy}
\label{sec:simres_batty}

\begin{figure}
\begin{center}
\begin{minipage}[b]{0.23\linewidth}
\centering
\includegraphics[width=\textwidth]{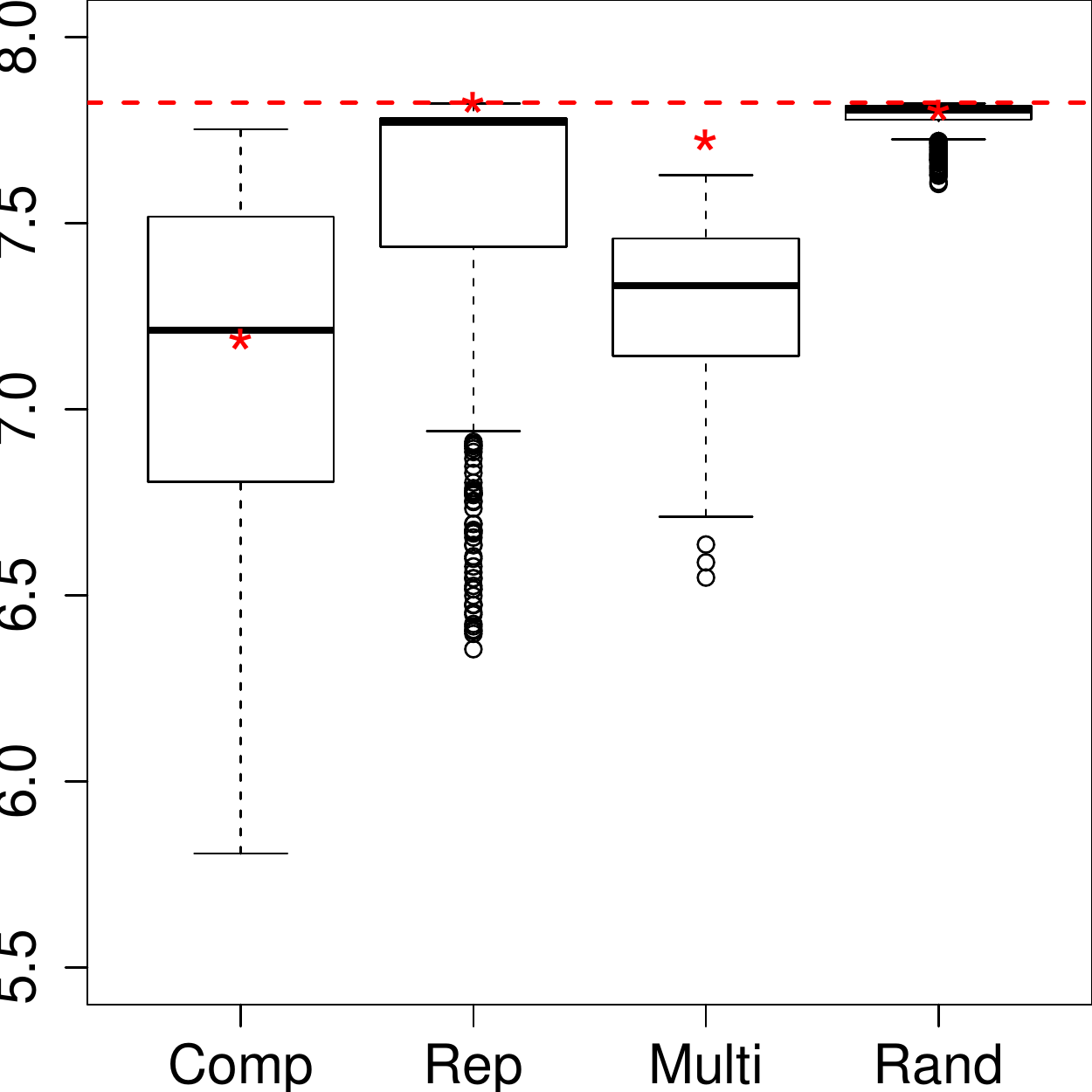}
\end{minipage}
\hspace{0.5cm}
\begin{minipage}[b]{0.69\linewidth}
\centering
\includegraphics[width=\textwidth]{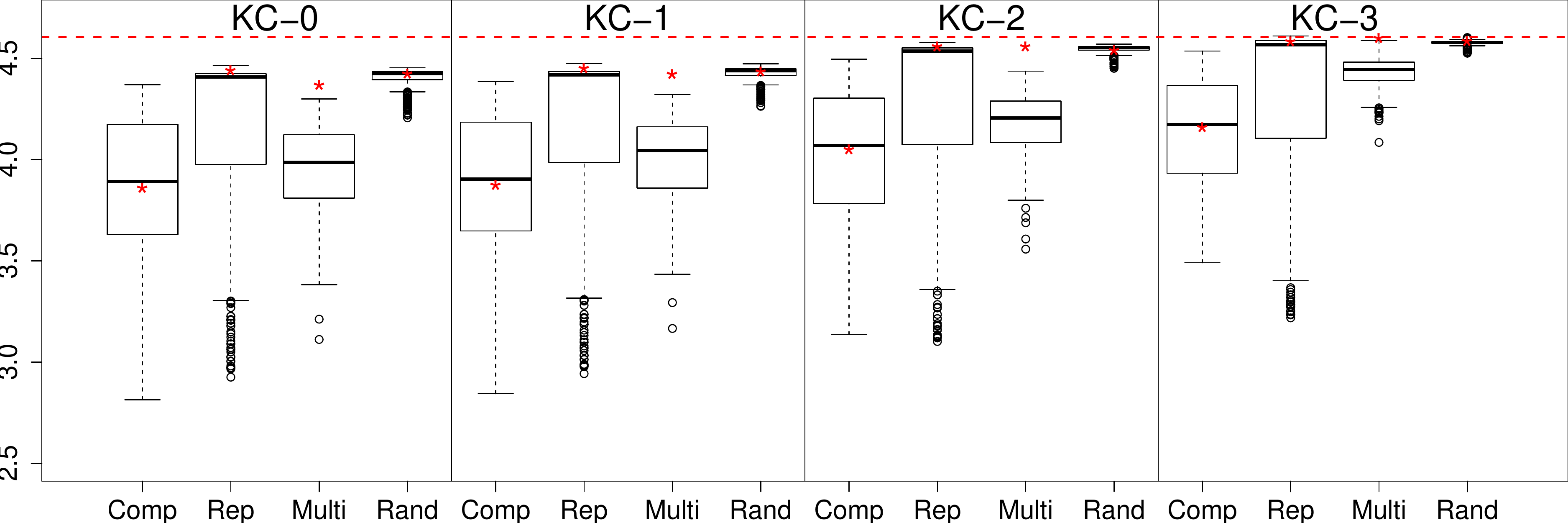}
\end{minipage}
\caption{Left panel: Batty's entropy, 1000 simulations. Right panel: \citeauthor{karlstrom}'s entropy, 1000 simulations. Each star identifies the entropy value computed on a uniformly distributed $X$. }
\label{fig:batty_X2}
\end{center}
\end{figure}

Results for Batty's entropy (\ref{eq:spaten}) are shown in Figure \ref{fig:batty_X2}, left panel, for the four spatial configurations described in Section \ref{sec:simdesign}. The dashed line corresponds to the index maximum $\log(T)=7.82$, which may only be reached with a repulsive or random spatial configuration. Batty's entropy measure, which does not make use of an adjacency matrix, is really able to detect a departure from a random configuration only when clustering occurs: the entropy distributions corresponding to compact and multicluster datasets are set on lower values than the distribution corresponding to random datasets. Conversely, the majority of the repulsive datasets generates entropy values that cannot be distinguished from those coming from random datasets. 
\citeauthor{karlstrom}'s entropy (\ref{eq:karl}) modifies Batty's entropy by including information about the neighbouring areas via an adjacency matrix and neglecting the sizes $T_g$. The index maximum is $\log(G)=4.61$.  When $d_0=0$ (only 1 neighbour for each area, i.e. the area itself) a special case of Batty's entropy without the $T_g$ terms is obtained. Figure \ref{fig:batty_X2}, right panel shows \citeauthor{karlstrom}'s entropies for the 4 neighbourhood options: \textit{KC-0} denotes the entropy measure computed using $d_0=0$, \textit{KC-1} using $d_1=2$, \textit{KC-2} using $d_2=5$ and \textit{KC-3} using $d_3=10$. As stated in Section \ref{sec:kc}, as the neighbourhood gets smaller, \citeauthor{karlstrom}'s entropy measures tend to the version without the $T_g$ terms of Batty's entropy (i.e. \textit{KC-0}). Results are very similar to Batty's ones, both for the 1000 generated datasets and for the special case of the uniform distribution of $X$, though the inclusion of the neighbourhood produces a monotone increase in all entropy values. The multicluster pattern is the most influenced by extending the neighbourhood: as the neighbourhood becomes larger, its entropy distribution gets closer and closer to the result of a random configuration. In this latter case, indeed, neighbourhood plays an important role since, under the random spatial configuration, areas present different spatial behaviours. Conversely, when areas tend to be similar, the inclusion of the neighbourhood does not substantially modify the conclusions.

\subsubsection{Entropy measures based on $Z$}
\label{sec:simres_Z}

\begin{figure}
\centering
    \includegraphics[width=.75\textwidth, height=5cm]{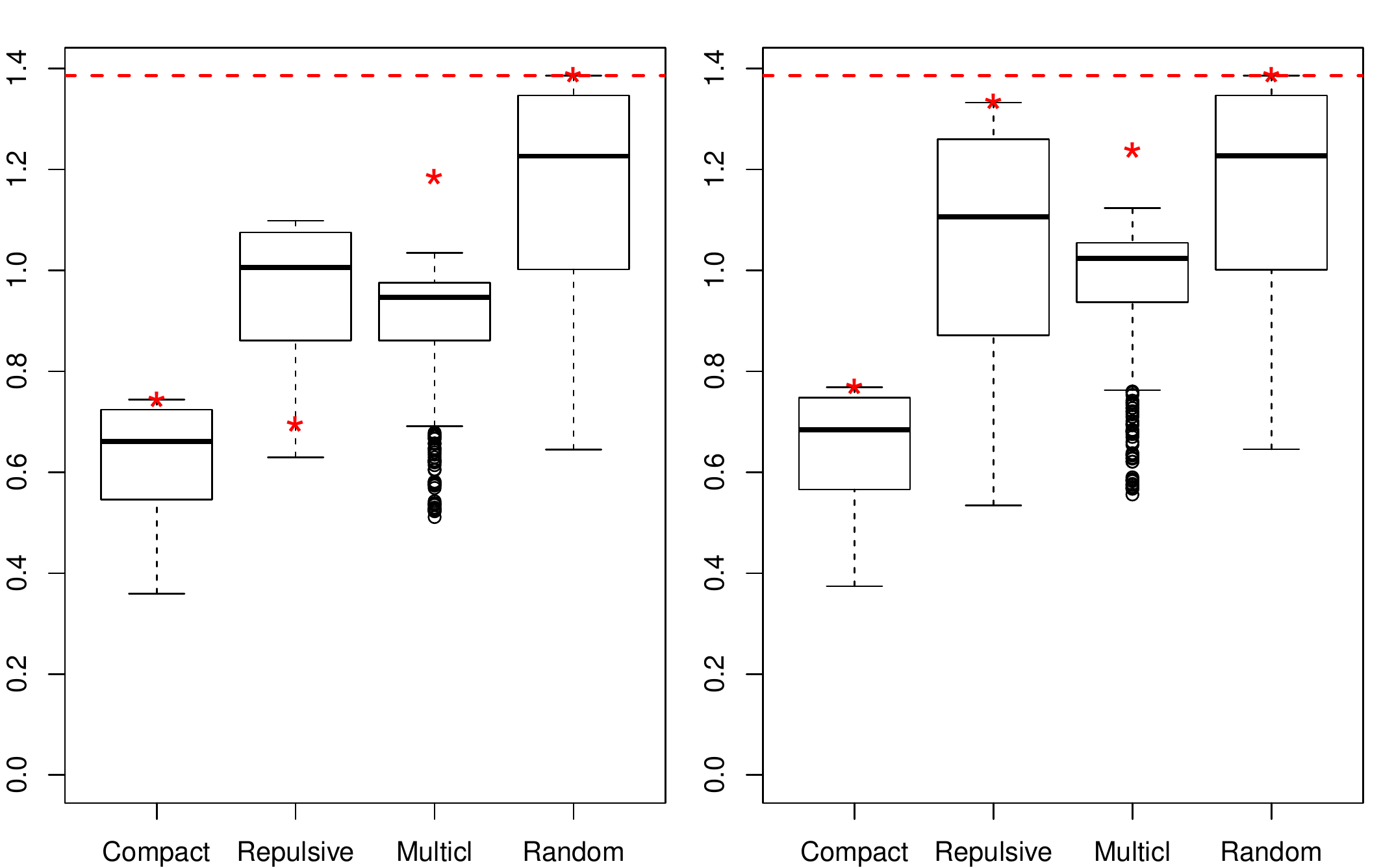}
          \caption{O'Neill's entropy (left) and Leibovici's entropy with $d=2$ (right), 1000 simulations. The dashed line corresponds to the index maximum; each star identifies the entropy value computed on a uniformly distributed $X$. }\label{fig:oneillX2}
\end{figure}

Results for O'Neill's spatial entropy (\ref{eq:oneill}) and Leibovici's spatial entropy (\ref{eq:leib}) at distance $d=2$ are displayed in Figure \ref{fig:oneillX2}; O'Neill's entropy is a special case of Leibovici's entropy with $d=1$. 
The main limit of the measures shown in Figure \ref{fig:oneillX2} is that for the repulsive patterns they produce on average much higher entropy values than for the compact ones. As stated in Section \ref{sec:reviewspace}, a proper spatial entropy measure accounts for the presence of a spatial pattern, but does not distinguish a negative correlation from a positive one. Therefore, results for the compact and repulsive pattern should be more similar than they appear here. This difference is mainly due to order preservation in building couples, which is also the reason why in the special case of uniform distribution for $X$ (star of the second boxplot in Figure \ref{fig:oneillX2}, left panel), the entropy value for the repulsive pattern is low but cannot reach the minimum value 0. For Leibovici's entropy, values for the repulsive patterns are even higher than in O'Neill's entropy, and nearly identical to those of the random patterns. This happens because different types of couples (second-order neighbours along the two cardinal directions, plus diagonals) are counted, increasing heterogeneity among couples. As for the compact and the multicluster data, the two indices behave the same way and return the same amount of information. This states that the choice of $d$ barely influences the entropy values, as long as $d$ is smaller than the cluster size. The more a pattern is compact, the stronger the expectation about the next $Z$ outcome, therefore, the compact configuration witnesses a low degree of surprise, whereas in the multicluster patterns the degree of surprise is higher. 
For the first three spatial configurations, the index maximum (dashed lines in Figure \ref{fig:oneillX2}) cannot be reached, because the spatial pattern does not allow the occurrence of the uniform distribution of $Z$. It may only be reached by a random pattern with a uniform distribution of $X$, where $H(Z|O)$ and $H(Z|L_2)$ are very similar to $H(Z)$. The random patterns are not influenced by space, therefore all measures lead to the same results irrespective of the distance. 

If the sign of O'Neill's entropy is changed, Parresol and Edwards' index (\ref{eq:gamma}) is obtained and contagion is measured instead of entropy. When (\ref{eq:gamma}) is also normalized, it evolves to the Relative Contagion index (\ref{eq:rc}); all above indices share the same basic idea, therefore comments to Figure \ref{fig:oneillX2} also hold for these indices.

\subsubsection{Spatial residual entropy and spatial mutual information}
\label{sec:simres_resid}

\begin{figure}
\centering
    \includegraphics[width=.8\textwidth]{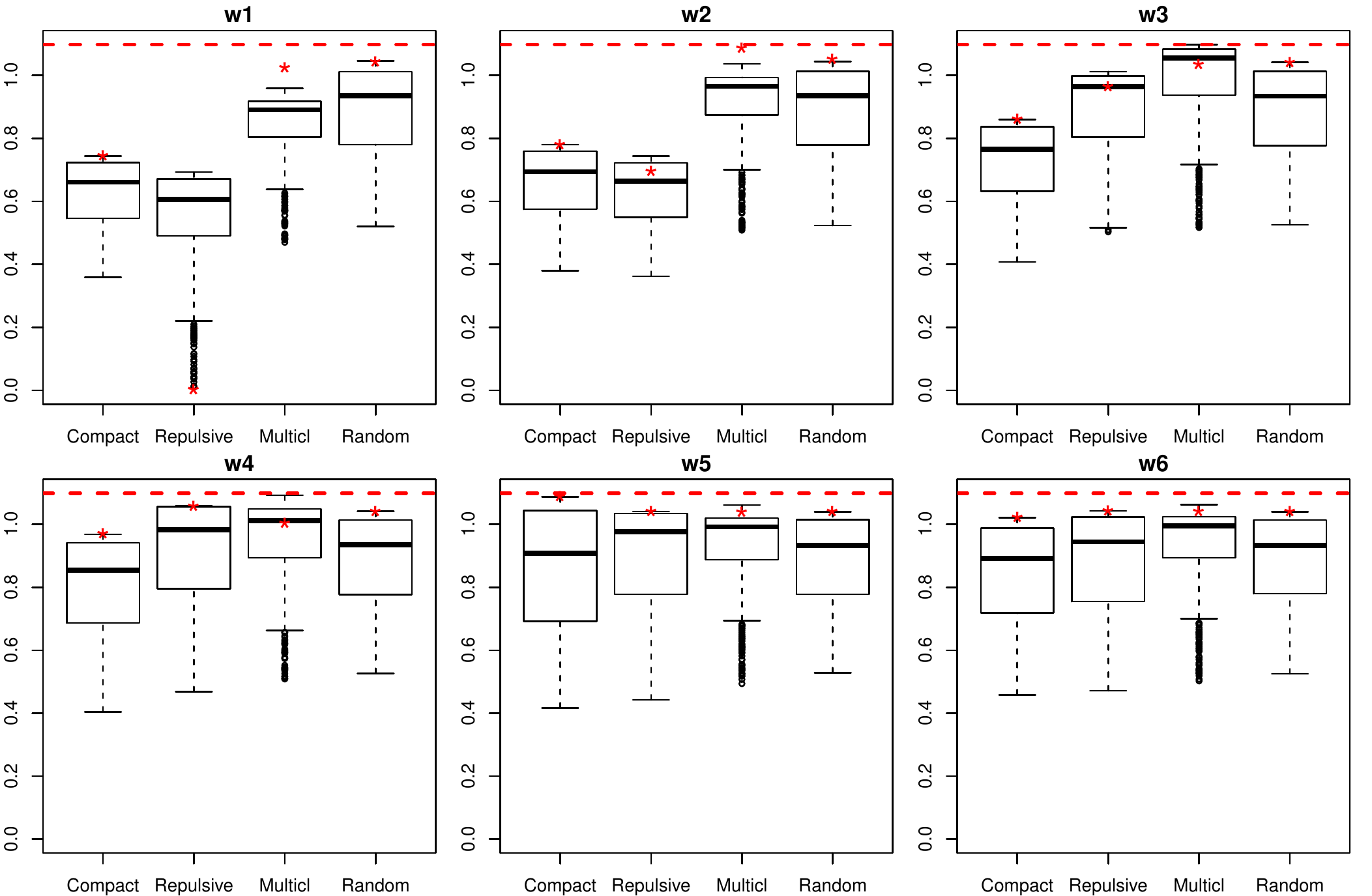}
          \caption{Spatial partial residual entropies, 1000 simulations. Each dashed line corresponds to the index maximum; each star identifies the entropy value computed on a uniformly distributed $X$. }\label{fig:respart_X2}
\end{figure}

Spatial partial residual entropies (\ref{eq:residZW_loc}) constitute the generalization of entropy measures based on $Z$ shown above, without order preservation. Figure \ref{fig:respart_X2} summarizes results for the partial terms at distances $w_1$ to $w_6$ (results for $w_7$, not reported here, are very similar to those for $w_6$). In the binary case, the panels referring to short distances, where spatial association occurs, are the most relevant. When further distances are taken into account and the ranges covered by the $W$ categories increase (as in the case of the 7 categories defined in Section \ref{sec:simdesign}), differences between the spatial configurations reduce.

In the first two panels of Figure \ref{fig:respart_X2} (i.e. at distances $w_1$ and $w_2$) interpreting the role of space is easier than by means of the entropy measures proposed in Sections \ref{sec:spatent}. Indeed, Batty's and \citeauthor{karlstrom}'s entropy only detect that space has a role in clustered patterns, while spatial residual entropy highlights a natural order across the four spatial configurations: the lower the spatial association, the higher the entropy.

The partial residual entropy values at distance $w_1$ in Figure \ref{fig:respart_X2} (where co-occurrences are couples of contiguous pixels) is O'Neill's entropy (\ref{eq:oneill}), reported in the left panel of Figure \ref{fig:oneillX2} without order preservation. Should the two partial residual entropies at distance $w_1$ and $w_2$ be summed, an unordered version of Leibovici's entropy, already shown in Figure \ref{fig:oneillX2} right panel, would be obtained. Since in the compact pattern most couples are formed by identical elements, order preservation is irrelevant with this configuration and results are very close to those reached by the previous entropy measures based on $Z$. A substantial improvemente is that the difference between entropies in compact and repulsive patterns is lower than in the case of O'Neill's and Leibovici's measures, while there is an evident difference between the situation of compact and repulsive patterns (the two strongly spatially associated ones) and the multicluster and random patterns (the two less spatially associated ones). This is a nice feature, since entropy measures should detect spatial association, irrespective of its type. Moreover, the entropy value for the uniform dataset with a repulsive pattern (star in the second boxplot of the first panel in Figure \ref{fig:respart_X2}) actually reaches the lower limit 0, since, when order is not preserved, all couples are of the same type in the perfect chessboard scheme. The entropy under a uniform distribution of $X$ at distance $w_2$ (star of the second boxplot of the second panel in Figure \ref{fig:respart_X2}) is larger than at distance $w_1$, as a greater number of couples is considered and couples formed by identical elements are also present. Entropy values are tendentially high for the multicluster dataset, closer to its maximum than in the case of O'Neill's and Leibovici's entropy. The random configuration entropy values are the highest, but they do not reach the index maximum as, since order is not preserved, a uniform distribution for $Z$ cannot be represented. Partial entropies allow to understand that the similarity between random and repulsive patterns in Leibovici's entropy (Figure \ref{fig:oneillX2}, right panel) is mainly due to what happens at distance $w_2$. This can be shown because partial terms (\ref{eq:residZW_loc}) consider different distance levels separately, while Leibovici's entropy counts all couples within a fixed distance without distinction.

As distances $w_3$ to $w_6$ are considered (third to sixth panel in Figure \ref{fig:respart_X2}), entropy values for the compact configuration increase slowly, while for the repulsive pattern they become more and more similar to those of the random pattern, as all $Z$ categories tend to be equally present. The multicluster configuration reaches the highest entropy values at distance categories $w_3$ and $w_4$. Entropy values for the random pattern remain similarly distributed across distances as expected, since no spatial association should be detected irrespective of the considered distances.

\begin{figure}
\centering
    \includegraphics[width=.4\textwidth, height=5cm]{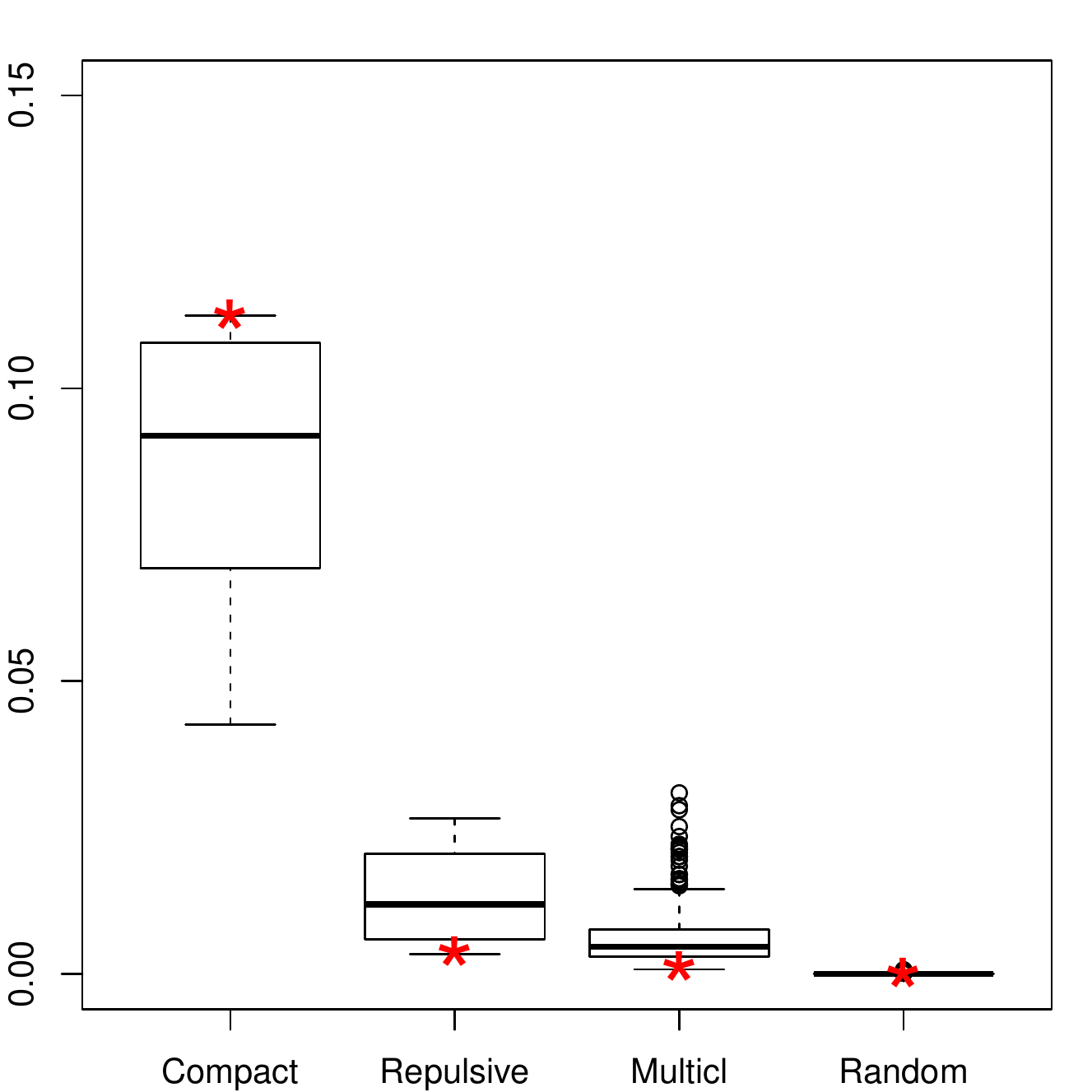}
          \caption{Proportional spatial mutual information, 1000 simulations. The stars identify the entropy value computed on a uniformly distributed $X$.}\label{fig:mutrel_X2}
\end{figure}

One focus of this work is on the contribution of the partial terms, rather than on the spatial global residual entropy (\ref{eq:residZW}), which, in accordance to the property of additivity, is a weighted sum of all partial terms (\ref{eq:residZW_loc}) for $w_k$, $k=1, \dots, 7$. For this reason, a graphical representation of the spatial global residual entropy is not shown here. Spatial global residual entropy (\ref{eq:residZW}) contributes to quantify the role of space: it allows to compute the spatial mutual information (\ref{eq:residZW_mut}), by subtracting (\ref{eq:residZW}) from $H(Z)$. The proportional version (\ref{eq:mutprop}) of the spatial mutual information is displayed in Figure \ref{fig:mutrel_X2}. 
Proportional spatial mutual information illustrates how the role of space decreases along the four considered spatial configurations: a globally appreciable influence of space is detected for the first two spatial patterns (compact and repulsive), the mutual information for the multicluster dataset is very small and no mutual information is detected over the random patterns, where no spatial structure is present and space does not help in explaining the data behaviour.
This measure also has the advantage of being easily interpretable: for instance, for the compact pattern, it says that nearly 10\% of the entropy is due to space (median of the first boxplot in Figure \ref{fig:mutrel_X2}). 

\begin{figure}
\centering
    \includegraphics[width=.9\textwidth]{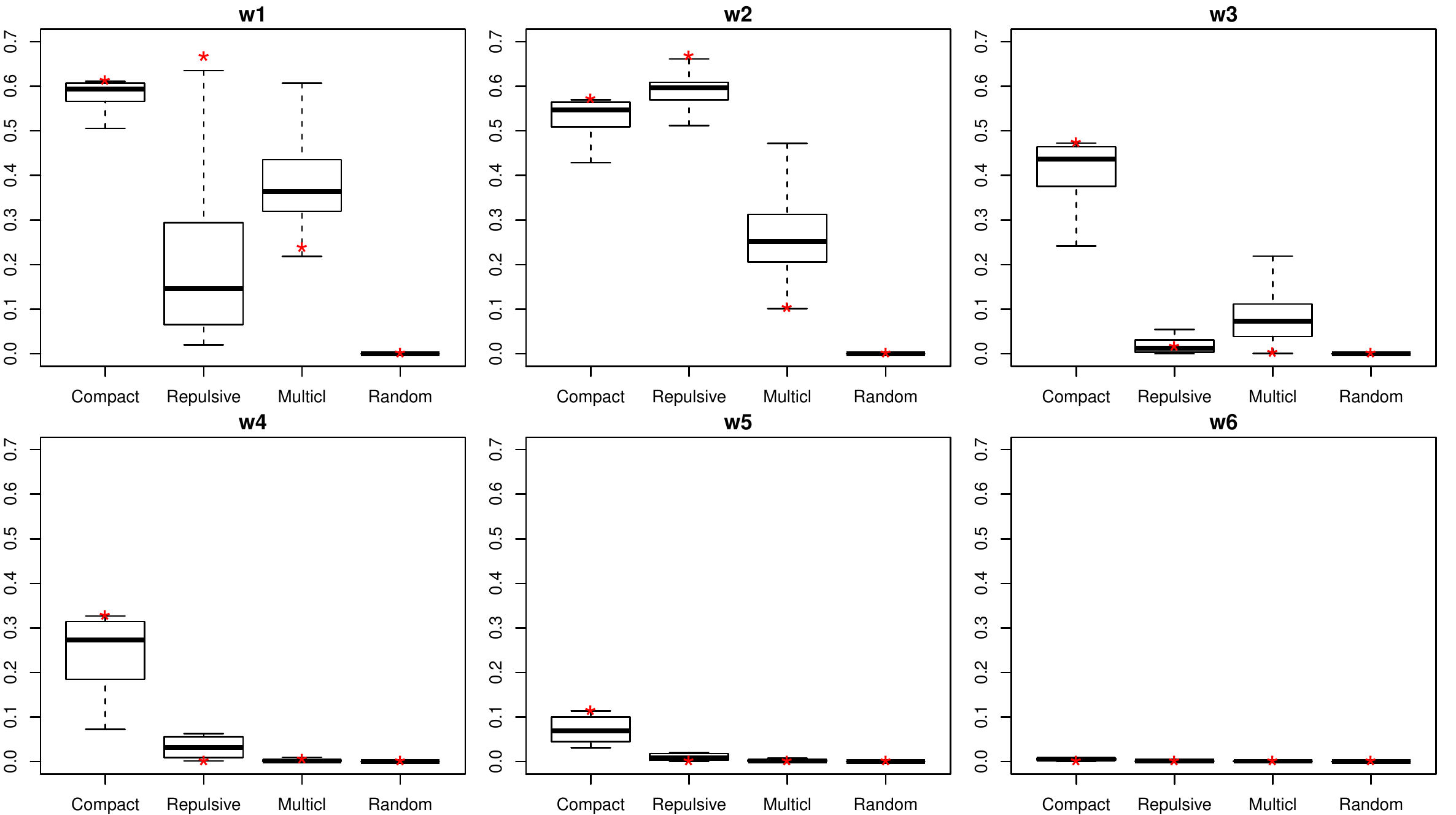}
          \caption{Spatial partial information, 1000 simulations. Each star identifies the entropy value computed on a uniformly distributed $X$.}\label{fig:partmut_X2}
\end{figure}

More detailed results are obtained by disaggregating the role of space at different distance categories:  spatial partial information terms (\ref{eq:partialterm_mut}) are shown in Figure \ref{fig:partmut_X2}. The spatial partial information constantly decreases for the compact patterns as distance increases, and behaves analogously with smaller values for the multicluster one. For the repulsive pattern, the spatial partial information takes high values for the first two distance ranges and drops from distance $w_3$ on. It is null at any distance range for the random patterns.

\subsection{Extension to data with more than two categories}
\label{sec:sim_resX5X20}

For $X5$ and $X20$, only the compact and random scenarios are investigated, since, as said in Section \ref{sec:simdesign}, when many unordered categories are present over a window, a repulsive or a multicluster pattern cannot be distinguished from a random one.

When switching from binary to data with more than two categories, all entropy values increase, since their maxima depend on the number of categories. Unnormalized indices are to prefer, as they account for diversity: the greater number of categories, the higher suprise/information about an outcome.

Irrespective of the chosen measure, all entropy values under a uniform distribution of $X$ (stars) are higher than the rest of the distribution (boxplots). This happens because, under the hypothesis of uniform distribution, all categories have the same importance. Conversely, when a random sequence of values from a multinomial distribution is generated, it does not always cover the whole range of potential categories. For instance, with $X20$, in several replicates less than 20 categories are actually produced. 

\begin{figure}
\centering
    \includegraphics[width=.7\textwidth]{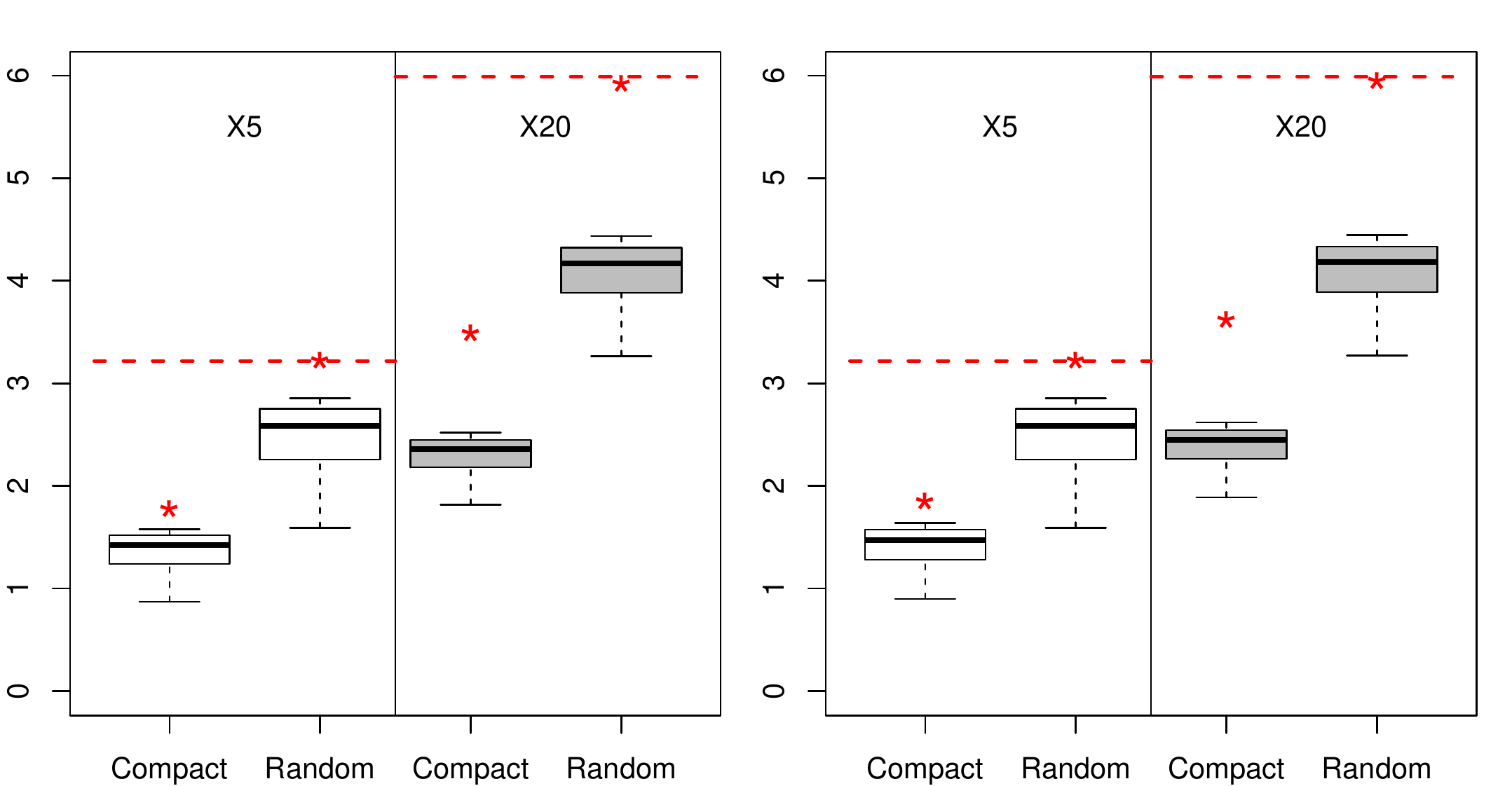}
          \caption{O'Neill's entropy (left) and Leibovici's entropy (right), 1000 simulations. Each dashed line corresponds to the index maximum; each star identifies the entropy value computed on a uniformly distributed $X$.
          }\label{fig:oneillX5X20}
\end{figure}

\begin{figure}
\centering
    \includegraphics[width=.9\textwidth]{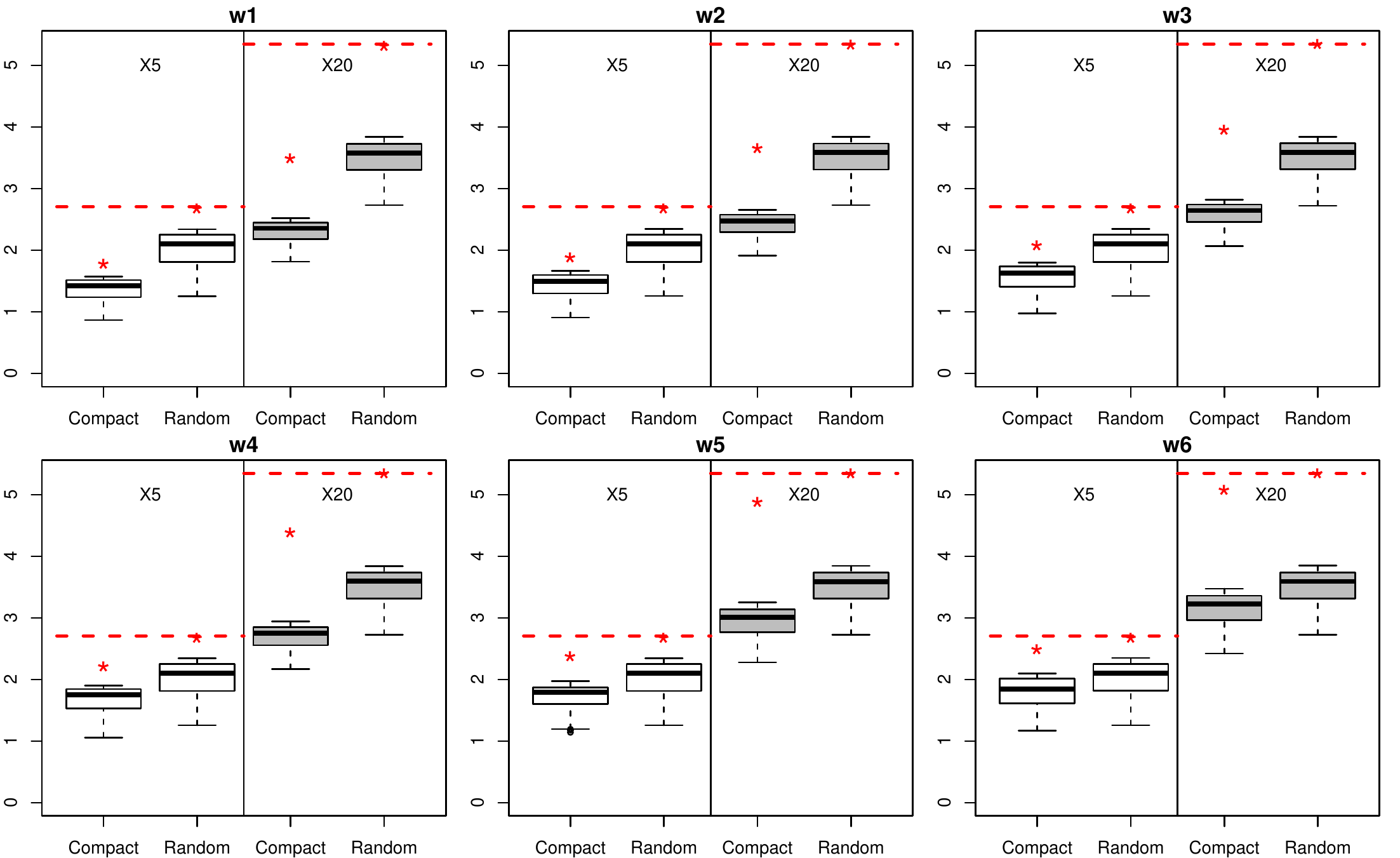}
          \caption{Spatial partial residual entropies, 1000 simulations. Each dashed line corresponds to the index maximum; each star identifies the entropy value computed on a uniformly distributed $X$. }\label{fig:respart_X5X20}
\end{figure}

All entropy measures computed for $X5$ and $X20$ are based on the trasformed variable $Z$; entropy maxima and number of categories may be retrieved in Table \ref{tab:Zcat}. O'Neill's entropy (\ref{eq:oneill}) and Leibovici's entropy (\ref{eq:leib}) are shown in Figure \ref{fig:oneillX5X20} and have similar behaviours: when the number of categories for $X$ increases, the distributions for the compact and random spatial configurations get farther apart.  
The partial terms (\ref{eq:residZW_loc}) of spatial residual entropy are shown in Figure \ref{fig:respart_X5X20}. As the distance category $w_k$ increases, the two spatial configurations lead to more similar entropy values; this is also due to the increasing range of the distance classes. On the other hand, as the number of categories increases (i.e. as the index maximum increases), the two distributions diverge, which is another desirable feature of the proposed measures.

\begin{figure}
\centering
    \includegraphics[width=.4\textwidth, height=5cm]{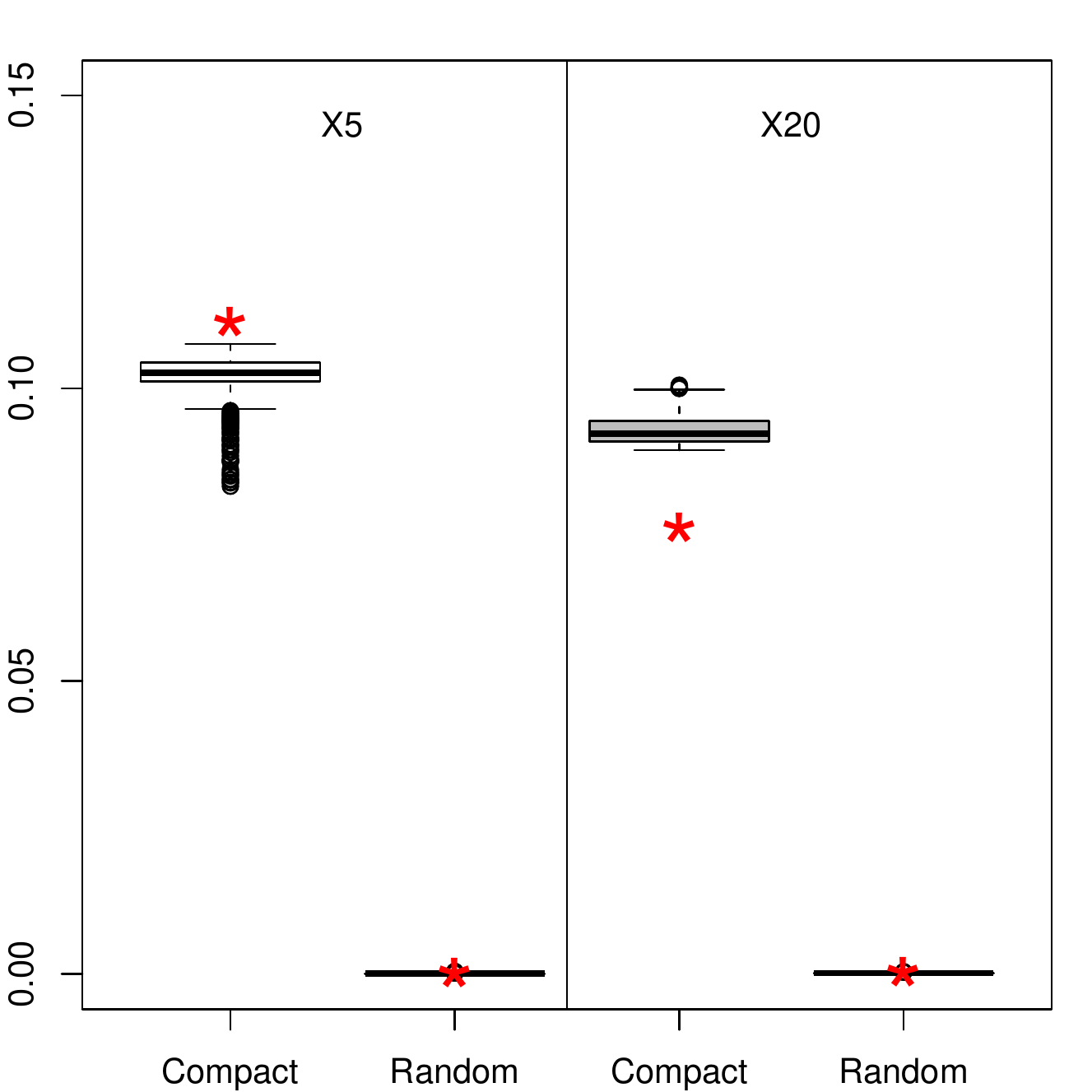}
          \caption{Proportional spatial mutual information, 1000 simulations. The stars identify the entropy value computed on a uniformly distributed $X$. }\label{fig:mutrel_X5X20}
\end{figure}

\begin{figure}
\centering
    \includegraphics[width=.9\textwidth]{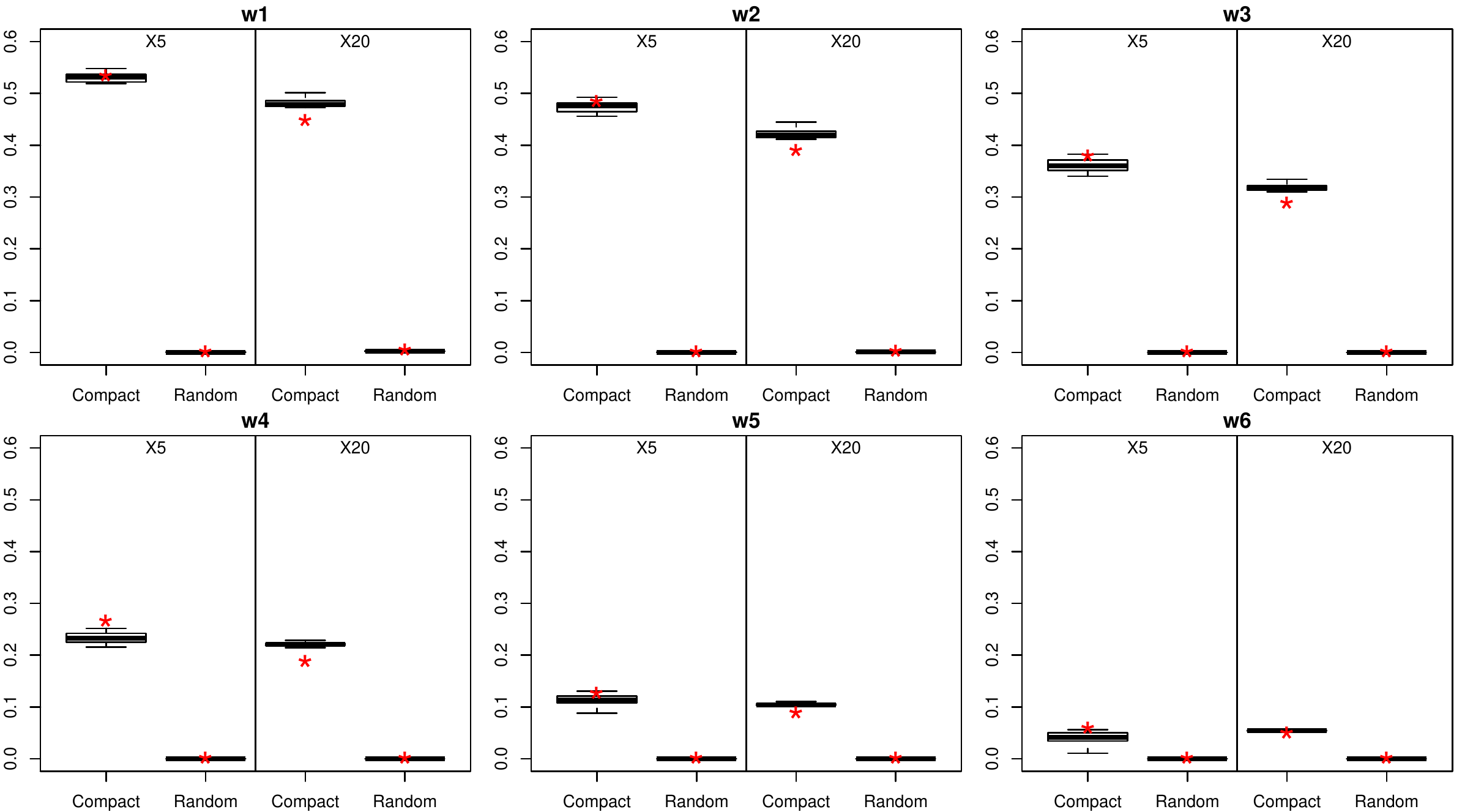}
          \caption{Spatial partial information, 1000 simulations. Each star identifies the entropy value computed on a uniformly distributed $X$.}\label{fig:partmut_X5X20}
\end{figure}

Proportional spatial mutual information is appreciable over both compact datasets (Figure \ref{fig:mutrel_X5X20}): including space as a variable returns information, meaning that the surprise of observing a certain outcome is reduced. For compact patterns, the distribution variability decreases when switching from $X2$ to $X5$ and $X20$ (Figures \ref{fig:mutrel_X2} and \ref{fig:mutrel_X5X20}); nevertheless, distributions are centered around similar values, therefore the role of space is constant across different numbers of categories. This is a key advantage of proportional spatial mutual information: the detected role of space is measured taking the number of categories into account. No mutual information is detected over the random patterns, where spatial structures are undetectable, and space does not help in explaining the data heterogeneity.
The same happens for all spatial partial information terms (Figure \ref{fig:partmut_X5X20}) referring to random patterns, irrespective of the number of categories. Rather, a monotone decrease occurs in the values obtained for compact patterns as distance increases. As the number of categories increases, spatial mutual information becomes less variable across generated data (lower interquartile ranges): the measure is very informative in distinguishing among different spatial configurations as regards the role of space.

\section{Discussion and conclusions}
\label{sec:disc}

When accounting for the role of space in determining the heterogeneity of the outcomes of the study variable, any type of spatial association, positive or negative, decreases entropy according to its strength. The sign of association is assessed by spatial correlation indices, which should not be confused with spatial entropy. 

The main innovation and merit of the two measures proposed in this paper are that they allow the entropy of a categorical variable to be decomposed into a term accounting for the role of space and a noise term summarizing the residual information. Results from the comparative study of Section \ref{sec:sim} show that the entropy measures proposed in this work, i.e. spatial residual entropy and spatial mutual information, are substantially different from the most popular indices. Their characteristics are summarized in Table \ref{tab:specchio}. Spatial residual entropy and spatial mutual information are the only measures that share all the listed desirable features.
\begin{table}
\caption{\label{tab:specchio}Characteristics of the entropy measures.}
\hskip-0cm\fbox{
\begin{tabular}{l|c|c|c|c|c|c}
     & Variable & Unord. & Info on $X$ cat. & Spatial & Additive & Bivariate\\
    \hline
    Shannon & $X$ & & $\blacksquare$ & & & \\
    Batty & $F$ & &  & $\blacksquare$ &  & \\
    \citeauthor{karlstrom} & $F$ & &  & $\blacksquare$ & $\blacksquare$ & \\
    O'Neill & $Z$ & & $\blacksquare$ & $\blacksquare$ & & \\
    Leibovici & $Z$ & & $\blacksquare$ & $\blacksquare$ & & \\
    Spatial residual entropy & \multirow{2}{*}{$Z,W$} & \multirow{2}{*}{$\blacksquare$} &
    \multirow{2}{*}{$\blacksquare$} & \multirow{2}{*}{$\blacksquare$} & \multirow{2}{*}{$\blacksquare$} & \multirow{2}{*}{$\blacksquare$}\\
    and mutual information &{}&{}&{}&{}&{}&{}\\
\end{tabular}}
\end{table}

First of all, the two proposed measures do not preserve the pair order, which is reasonable in spatial analysis. Indeed, spatial phenomena are not usually assumed to have a direction: the primary interest lies in understanding the spatial heterogeneity of data over a specific area, considering neighbourhood in any direction. Neglecting the order allows the presence of spatial patterns to be better distinguished from spatial randomness. 

Unnormalized entropy indices ought to be preferred, in agreement with \cite{parresol}, in order to distinguish between situations with different numbers of categories of $X$ and, consequently, of $Z$. Most spatial association measures, on the contrary, need normalization, since they ought not to depend on the number of data categories. Entropy is not primarily conceived to measure spatial association, rather it measures the surprise concerning an outcome, therefore, given a fixed degree of spatial association, the surprise has to be higher for a dataset with more categories. Normalized entropy measures may be only preferred in special cases to achieve easily interpretable results.

In addition, spatial residual entropy and spatial mutual information improve Batty's (\ref{eq:spaten}) and \citeauthor{karlstrom}'s (\ref{eq:karl}) entropies. They also enjoy \citeauthor{karlstrom}'s property of additivity and consider space, but, since they do not lose information about the variable categories, the two measures answer a wider set of questions. Unlike \citeauthor{karlstrom}'s measure that refers to a univariate approach basing on a unique adjacency matrix, the proposed measures consider different matrices to cover all the range of possible distances and exploit the bivariate properties to unify the partial results.

Moreover, spatial residual entropy and spatial mutual information constitute a substantial theoretical improvement with regard to O'Neill's entropy, Parresol and Edward's index and the Relative Contagion index, as they only consider contiguous pairs (at distance 1), while (\ref{eq:residZW}) and (\ref{eq:spatial_mutual}) give a global view of what happens over a dataset, since they also consider distances greater than 1.
Leibovici's entropy (\ref{eq:leib}) is a more general measure which extends to further distances. Compared to it, spatial residual entropy and spatial mutual information have additional advantages. First of all, they consider unordered couples with the aforementioned consequences. Secondly, Leibovici's measure does not allow any deeper inspection, whereas (\ref{eq:residZW}) and (\ref{eq:spatial_mutual}) can investigate what happens at different distance ranges. This enriches the interpretation of results, as knowledge is available about what distances are more important for the spatial association of a studied phenomenon. 

In the study presented here, the most interesting distances are the small ones. Real life situations are very challenging opportunities, where different spatial configurations can arise; at this regard, (\ref{eq:residZW}) and (\ref{eq:spatial_mutual}) are very flexible in identifying the most informative distance to properly interpret the phenomenon under study. Indeed, spatial residual entropy is able to detect this aspect by discerning the contribution to entropy of different distances through its partial versions (\ref{eq:residZW_loc}) which can be summarized to form the global (\ref{eq:residZW}) or further decomposed as wished. Thus, the categories $w_k$ must be suitably proposed according to the context, as the less interesting distances should be aggregated while the most interesting ones ought to be analyzed in detail. 

Furthermore, in spatial global residual entropy the definition of equal-sized distance classes is not required, since weights suitably resume the spatial partial residual entropies to properly form the global version. When the spatial global residual entropy (\ref{eq:residZW}) is computed, probabilities of couples that occur at different distances, $p(z_r|w_k)$, are weighted by $p(w_k)$, so that the relative weight of all distances is respected.

Lastly, spatial mutual information is a further tool to exploit for quantifying the overall information brought by the inclusion of space. It is different from zero when it is possible to recognize a (positive or negative) spatial pattern. It can be decomposed the same way as spatial global residual entropy to investigate the role of space at each distance range. In addition, its ability to detect the role of space is not influenced by the number of categories of $X$, as shown in Section \ref{sec:sim_resX5X20}.

This work provides a complete toolbox for analyzing spatial data where distance is believed to play a role in determining the heterogeneity of the outcomes. The first step of this analysis consists in computing Shannon's entropy of $Z$ to keep as a reference value. Secondly, spatial mutual information is computed and its proportional version identifies the overall role of space. According to this result, distance classes are then suitably defined in order to investigate the partial terms. In particular, partial information terms help to understand whether space plays a relevant role at each distance class, while spatial partial residual entropies focus on the heterogeneity of the study variable due to other sources. The comparison of partial terms across distances is also helpful to grasp the spatial behaviour of the study variable.

The proposed entropy measures may be extended to spatially continuous data presenting a finite number of categories, such as marked spatial point patterns.

\bigskip\noindent\textbf{Acknowledgements}\\
This work is developed under the PRIN2015 supported-project 'Environmental processes and human activities: capturing their interactions via statistical methods (EPHASTAT)' funded by MIUR (Italian Ministry of Education, University and Scientific Research).
\\
As regards author Linda Altieri, the research work underlying this paper was partially funded
by an FIRB 2012 [grant number RBFR12URQJ] 'Statistical modelling of environmental phenomena: pollution,
meteorology, health and their interactions' for research projects by the Italian Ministry of Education, Universities and
Research.

\nocite{*}
\bibliographystyle{chicago}
\bibliography{bibdatabase_entropy}

\end{document}